\documentclass[journal=jacsat,manuscript=article]{achemso}

\usepackage{adjustbox}
\usepackage[colorlinks=true, allcolors=blue]{hyperref}
\usepackage{amsmath}
\usepackage{kmath}
\usepackage{amssymb}
\usepackage[version=4]{mhchem}
\usepackage{bbding}
\usepackage{setspace}
\usepackage{placeins}
\usepackage{pifont}
\usepackage{xcolor}
\usepackage[normalem]{ulem}
\usepackage{verbatim}
\usepackage{indentfirst}
\usepackage{tabularray}
\usepackage{textgreek}
\newcommand{\didv}{{d}\textit{I}/{d}\textit{V} }

\author{Jeong Ha Hwang}
{\altaffiliation{These authors contributed equally.}
\affiliation[1]{nanotech@surfaces Laboratory, Empa - Swiss Federal Laboratories for Materials Science and Technology, 8600 D\"ubendorf, Switzerland}
\alsoaffiliation[2]{Department of Information Technology and Electrical Engineering, ETH Z\"urich, 8092 Z\"urich, Switzerland}
}

\author{Nicol\`o Bassi}
\altaffiliation{These authors contributed equally.}
\affiliation[1]{nanotech@surfaces Laboratory, Empa - Swiss Federal Laboratories for Materials Science and Technology, 8600 D\"ubendorf, Switzerland}

\author{Mayada Fadel}
\affiliation[3]{Department of Materials Science and Engineering, Rensselaer Polytechnic Institute, Troy, NY 12180, USA}

\author{Oliver Braun}
\affiliation[4]{Transport at Nanoscale Interfaces Laboratory, Empa - Swiss Federal Laboratories for Materials Science and Technology, 8600 D\"ubendorf, Switzerland}
\alsoaffiliation[5]{Department of Physics, University of Basel, 4056 Basel, Switzerland}

\author{Tim Dumslaff}
\affiliation[6]{Max Planck Institute for Polymer Research, 55128 Mainz, Germany}

\author{Carlo Pignedoli}
\affiliation[1]{nanotech@surfaces Laboratory, Empa - Swiss Federal Laboratories for Materials Science and Technology, 8600 D\"ubendorf, Switzerland}

\author{Michael Stiefel}
\affiliation[4]{Transport at Nanoscale Interfaces Laboratory, Empa - Swiss Federal Laboratories for Materials Science and Technology, 8600 D\"ubendorf, Switzerland}

\author{Roman Furrer}
\affiliation[4]{Transport at Nanoscale Interfaces Laboratory, Empa - Swiss Federal Laboratories for Materials Science and Technology, 8600 D\"ubendorf, Switzerland}

\author{Hironobu Hayashi}
\affiliation[7]{Center for Basic Research on Materials, National Institute for Materials Science (NIMS), 1-2-1 Sengen, Tsukuba, Ibaraki 305-0047, Japan}

\author{Hiroko Yamada}
\affiliation[8]{Kyoto University, Kyoto, 606-8501, Japan}

\author{Akimitsu Narita}
\affiliation[6]{Max Planck Institute for Polymer Research, 55128 Mainz, Germany}

\author{Klaus M\"ullen}
\affiliation[6]{Max Planck Institute for Polymer Research, 55128 Mainz, Germany}

\author{Michel Calame}
\affiliation[4]{Transport at Nanoscale Interfaces Laboratory, Empa - Swiss Federal Laboratories for Materials Science and Technology, 8600 D\"ubendorf, Switzerland}
\alsoaffiliation[5]{Department of Physics, University of Basel, 4056 Basel, Switzerland}
\alsoaffiliation[12]{Swiss Nanoscience Institute, University of Basel, 4056 Basel, Switzerland}

\author{Mickael L. Perrin}
\affiliation[2]{Department of Information Technology and Electrical Engineering, ETH Z\"urich, 8092 Z\"urich, Switzerland}
\alsoaffiliation[4]{Transport at Nanoscale Interfaces Laboratory, Empa - Swiss Federal Laboratories for Materials Science and Technology, 8600 D\"ubendorf, Switzerland}
\alsoaffiliation[9]{Quantum Center, ETH Zürich, 8093 Zürich, Switzerland}

\author{Roman Fasel}
\affiliation[1]{nanotech@surfaces Laboratory, Empa - Swiss Federal Laboratories for Materials Science and Technology, 8600 D\"ubendorf, Switzerland}
\alsoaffiliation[11]{Department of Chemistry, Biochemistry and Pharmaceutical Sciences, University of Bern, 3012 Bern, Switzerland}

\author{Pascal Ruffieux}
\affiliation[1]{nanotech@surfaces Laboratory, Empa - Swiss Federal Laboratories for Materials Science and Technology, 8600 D\"ubendorf, Switzerland}

\author{Vincent Meunier}
\affiliation[10]{Department of Engineering Science and Mechanics, Pennsylvania State University, University Park, PA 16802, USA}

\author{Gabriela Borin Barin}
\affiliation[1]{nanotech@surfaces Laboratory, Empa - Swiss Federal Laboratories for Materials Science and Technology, 8600 D\"ubendorf, Switzerland}

\email{gabriela.borin-barin@empa.ch}

\title{Optimized Synthesis and Device Integration of Long 17-Atom-Wide Armchair Graphene Nanoribbons}

\begin{document}

\
\begin{abstract}
Seventeen-carbon-atom-wide armchair graphene nanoribbons (17-AGNRs) are promising candidates for high-performance electronic devices due to their narrow electronic bandgap. Atomic precision in edge structure and width control is achieved through a bottom-up on-surface synthesis (OSS) approach from tailored molecular precursors in ultra-high vacuum (UHV). This synthetic protocol must be optimized to meet the structural requirements for device integration, with ribbon length being the most critical parameter. Here, we report optimized OSS conditions that produce 17-AGNRs with an average length of approximately $\sim$17~nm. This length enhancement is achieved through a gradual temperature ramping during an extended annealing period, combined with a template-like effect driven by monomer assembly at high surface coverage. The resulting 17-AGNRs are comprehensively characterized in UHV using scanning probe techniques and Raman spectroscopy. Raman measurements following substrate transfer enabled the characterization of GNRs' length distribution on the device substrate and confirmed their stability under ambient conditions and harsh chemical environments, including acid vapors and etchants. The increased length and ambient stability of the 17-AGNRs lead to their reliable integration into device architectures.  As a proof of concept, we integrate 17-AGNRs into field-effect transistors (FET) with graphene electrodes and confirm that electronic transport occurs through the GNRs. This work demonstrates the feasibility of integrating narrow-bandgap GNRs into functional devices and contributes to advancing the development of carbon-based nanoelectronics.
\end{abstract}

\clearpage

\section{Introduction}
Graphene nanoribbons (GNRs) are narrow, one-dimensional strips of graphene that exhibit unique physical properties compared to their two-dimensional counterpart. A key characteristic of GNRs is the emergence of a bandgap due to quantum confinement in one dimension, making them promising candidates for nanoelectronic applications.~\cite{Nakada1996Edge,Zhou2014Modulating, Son2006Energy, Lee2005Magnetic} Their electronic, magnetic, and optical properties can be further tailored through structural modifications, such as variations in edge geometry and ribbon width. 

Depending on their edge geometry, GNRs can be classified into different groups, such as armchair (AGNR)\cite{cai2010atomically,kimouche2015ultra}, zigzag (ZGNR)\cite{ruffieux2016surface,blackwell2021spin}, chevron (CGNR)\cite{cai2010atomically,teeter2019surface}, and chiral~\cite{li2021topological}. Among these, AGNRs have been the most extensively studied, both theoretically and experimentally, due to their width-tunable electronic properties and ambient stability, which make them suitable for integration into switching devices.\cite{mutlu2021short}  AGNRs are further classified into three families based on the number of carbon atoms across their width: $3p$, $3p + 1$, and $3p + 2$, where $p$ is an integer. Within each family, the bandgap decreases monotonically with increasing width. However, because an atom-by-atom width increase alternates between these families, the overall bandgap variation is non-monotonic.~\cite{yang2007quasiparticle,Overbeck2019Universal,Son2006Energy,Saraswat2021Materials} AGNRs belonging to the $3p + 2$ family have the narrowest electronic bandgap of these three families, which can result in more transparent Schottky barriers at the metal contact, thereby reducing contact resistance and improving device performance. To date, 5-, 7-, 9-, and 13-AGNRs have been successfully integrated into field effect transistor (FET) devices~\cite{bennett2013bottom,Llinas2017Short,Jacobberger2017High,Ohtomo2018Graphene,Zhang2023Contacting,mutlu2021short,zhang2023tunable,huang2023edge}, with 5-AGNRs being the only representative of the $3p + 2$ family.~\cite{Borin2022Growth,El2020Controlled}

Due to the strong correlation between GNR geometry and electronic properties, synthesizing defect-free AGNRs with atomic-level precision is crucial for the development of devices with high reproducibility. On-surface synthesis (OSS) has emerged as a robust and reproducible bottom-up technique for achieving such control.~\cite{cai2010atomically,talirz2017surface,houtsma2021atomically} This approach relies on surface-assisted activation of specially designed organic precursors, which undergo polymerization via Ullmann-type coupling, followed by a cyclodehydrogenation reaction that planarizes the polymers, forming the final GNR structure.

Among the AGNRs accessible via OSS, 17-AGNRs represent a compelling candidate for device integration owing to their low electronic bandgap.~\cite{Yamaguchi2020Small} Density functional theory (DFT) predicts a bandgap of 0.14~eV, while experimental measurements on Au(111) by scanning tunneling spectroscopy indicate a gap of ~0.19~eV.~\cite {Yamaguchi2020Small} While such a low bandgap could lead to lower resistance at the metal contacts, it also raises concerns regarding the ribbons’ chemical stability and suitability for substrate transfer and device integration.~\cite{Borin2019Surface,Borin2015Optimized}. Moreover, based on standard electrode patterning techniques such as e-beam lithography, GNRs must be 15~nm or longer to bridge the source and drain with an effective contact area.~\cite{Braun2021Optimized}. 

In this work, we present an optimized growth protocol for synthesizing long and high-quality 17-AGNRs. By employing a high precursor coverage (near monolayer) and a gradual annealing process, we minimize defects and promote ribbon elongation. Scanning tunneling microscopy (STM) confirms that our approach yields 17-AGNRs with an average length of approximately 17~nm, while scanning tunneling spectroscopy (STS) reveals a bandgap of 0.45~eV. We perform Raman spectroscopy measurements to monitor the structural quality and ambient stability of the GNRs throughout their fabrication process: from growth in ultra-high vacuum to \textit{ex-situ} transfer into the devices. We further conduct a combined experimental and theoretical Raman study to identify characteristic vibrational fingerprints of 17-AGNRs and assess their length distribution.  Lastly, leveraging the improved structural quality and increased length of the ribbons, we fabricate and electrically characterize FET devices incorporating 17-AGNRs. 

\section{Results \& discussion}
\subsection{Synthesis and length optimization of 17-AGNR}

\begin{figure}[!htbp]%
\centering
\includegraphics[width=0.9\textwidth]{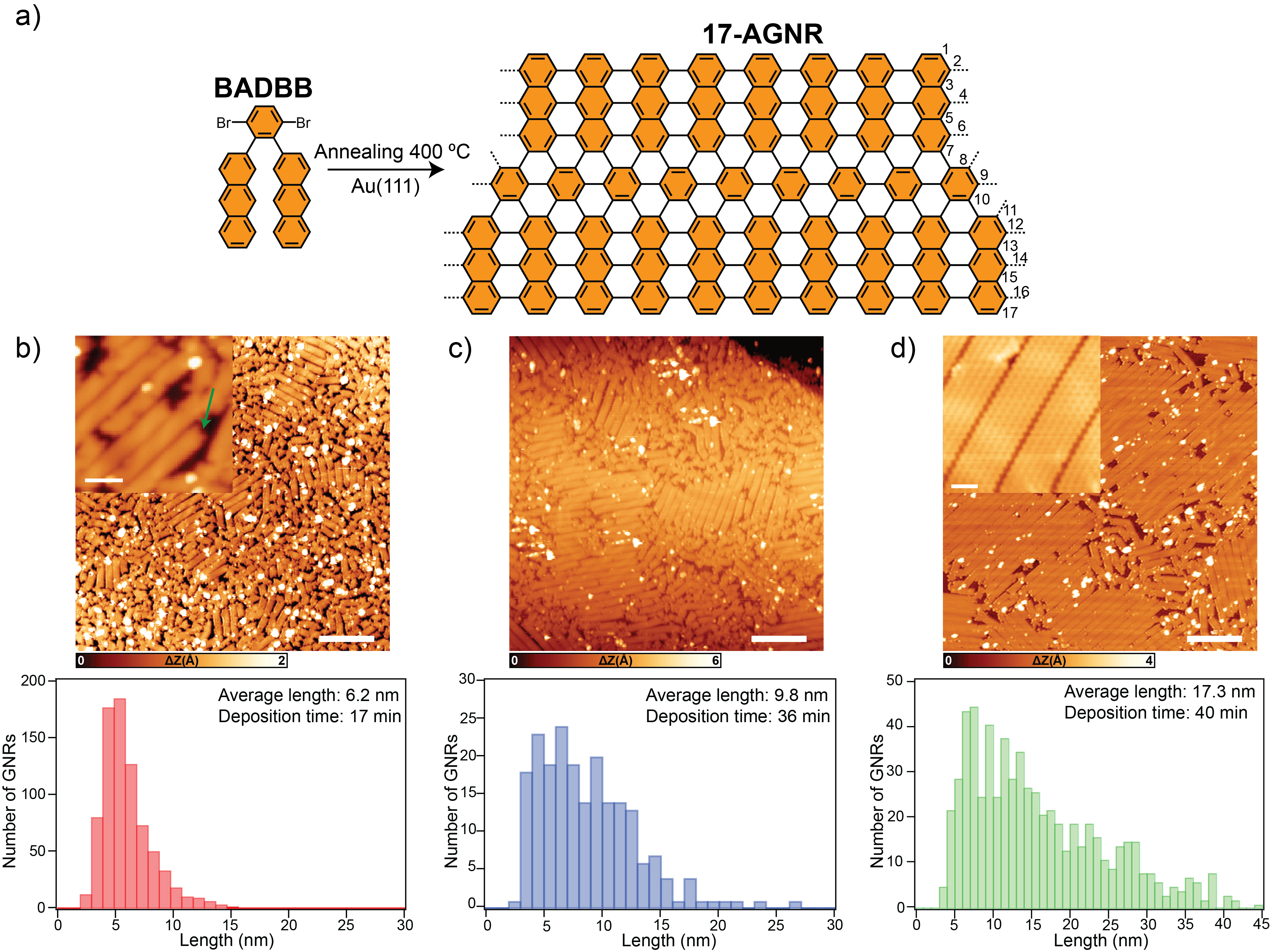}
\caption{ \textbf{On-surface synthesis and length optimization of 17-AGNR}. a) Chemical model of  \textbf{BADBB} and final \textbf{17-AGNR}. b) STM image of 17-AGNRs after slow annealing ramp (deposition time: 17~min; scale bar: 10~nm; scanning parameters: -1.5~V and 30~pA). Inset: high resolution image showing the ends of the 17-AGNRs unit (scale bar: 10~nm scanning parameters: -1.0~V and 50~pA). The ribbons grow short (length distribution histogram at the bottom). c) STM image of 17-AGNR using higher initial coverage (deposition time: 36~min; scale bar: 10~nm; scanning parameters: -1.5~V and 30~pA). The ribbons grow longer with an average length of 9.8~nm. d) STM image of 17-AGNR using higher initial coverage ( scale bar: 10~nm scanning parameters: -1.5~V and 30~pA). Inset: zoom-in image of 17-AGNR units with the expected width of 2.48~nm (scale bar: 1~nm). The ribbons grow longer than previous examples (average length of 17.3~nm with single ribbons exceeding 50~nm). STM measurements reported here have been performed at room temperature.}
\label{f1}
\end{figure}

The synthesis and length optimization of  17-AGNRs were performed in ultra-high vacuum (UHV) and characterized by scanning tunneling microscopy (STM) (Figure~\ref{f1}a). The molecular precursor, \textbf{BADBB} (1,2-bis-(anthracenyl)-3,6-dibromobenzene), features a central benzene ring bonded to two anthracene units, which ultimately form the 17-carbon-wide GNR. \textbf{BADBB} was synthesized \textit{via} Suzuki coupling of 1,4-dibromo-2,3-diiodobenzene and anthracen-2-ylboronic acid as previously reported~\cite{Yamaguchi2020Small,dumslaff_2017} and subsequently sublimated at 250~°C onto a clean Au(111) substrate in UHV at room temperature.
A large-scale STM topography image of the surface after precursor deposition \textbf{BADBB} reveals approximately 70$\%$ of a monolayer coverage (figure~\ref{fS_BADBB-on-Au}). The non-planar \textbf{BADBB} molecules self-assemble in large islands with different domain orientations. Within these assemblies (Figure~\ref{fS_BADBB-on-Au}b), the molecules adopt distinct geometries: a linear arrangement (with two opposite orientations) and an intermediate configuration (highlighted in green, blue, and yellow, respectively). At 77~K, STM lacks sufficient resolution due to molecular mobility, requiring support from DFT simulations to discern the absorption configurations. Figure~\ref{fS_BADBB-on-Au}c shows the relaxed geometry of a single \textbf{BADBB} molecule and the corresponding STM simulation (left and right panels, respectively). Because of steric hindrance between the side groups, the anthracene units do not lie flat on the surface but adopt a tilted configuration.

\FloatBarrier

After deposition on Au(111), the samples were gradually annealed to induce, first, polymerization step and subsequently cyclodehydrogenation reactions, yielding 17-AGNRs. The coupling of \textbf{BADBB} radicals with confined movement on the growth substrate requires a 180° rotation between adjacent monomers, analogous to 9-AGNRs precursors~\cite{Talirz2016OnSurface}. Dehalogenation of bromine-containing precursors typically begins at 150~°C~\cite{di2018surface, cai2010atomically}, while cyclodehydrogenation—where H atoms are released to form C–C bonds— initiates above 350~°C. Longer GNRs are expected to form when these two processes are well separated, preventing polymer passivation by hydrogen atoms released during the cyclodehydrogenation step. To achieve this, we implemented a very slow temperature ramp of 2~°C/min from 150~°C to 400~°C, ensuring a clear separation between dehalogenation and cyclodehydrogenation steps.

Initial synthesis yielded relatively short GNRs averaging 6.2~nm in length, with unreacted fragments scattered across the surface (Figure~\ref{f1}b). To further promote the growth of longer GNRs, we increased the precursor deposition time to raise the initial \textbf{BADBB} coverage~\cite{ishii2020quality}. As shown in figures~\ref{f1}c and d, this led to average GNR lengths of 9.8~nm and 17.3~nm, respectively. The longest GNRs were obtained when they grew in domains where they self-organized into parallel, ordered islands (Figure~\ref{f1}d)). This templating effect is known to enhance ribbon growth~\cite{moreno2018surface}. Moreover, higher coverage reduced the number of molecular fragments between GNRs, a possible cause of short units. The inset of Figure~\ref{f1}d shows a zoomed-in image confirming the expected width of 2.45~nm.~\cite{Yamaguchi2020Small}

Although the \textbf{BADBB} design suggests staggered termini, most 17-AGNRs instead exhibit two short protrusions at approximately 90° angles (green arrows in the insets of Figure~\ref{f1}b and~\ref{fS_End-segment}a). To investigate these terminations, we prepared a low-coverage sample,  yielding short GNRs where both termini exhibit the aforementioned protrusions (Figure~\ref{fS_End-segment}b). DFT calculations (\ref{fS_End-segment}c) reveal that these arise from the rotation of the anthracene unit in the last molecular precursor segment and the subsequent formation of pentagons. This interpretation is supported by the close agreement between the experimental and simulated STM images in panels b and d in figure~\ref{fS_End-segment}, respectively. Once the anthracene units rotate and C-C bonds form, further precursor attachment is sterically hindered, effectively terminating GNR growth. This suggests that the longest 17-AGNRs are achieved under high precursor coverage and a slow annealing ramp, where the flipping of the anthracene units is strongly reduced.

For switching device applications, the electronic bandgap of GNRs is a critical parameter. Due to their large width, 17-AGNRs are expected to have smaller bandgaps than narrower AGNRs in the $3p + 2$ family. DFT-PBE (Perdew-Burke-Ernzerhof) calculations (see Methods) for an infinitely long GNR predict a gas-phase bandgap of roughly 0.14~eV, significantly lower than for 7-AGNRs (1.58~eV) and 9-AGNRs (0.79~eV). While DFT underestimates absolute values, experimental bandgaps on Au(111) are reported as 2.3~eV for 7-AGNRs~\cite{ruffieux2012electronic} and 1.4~eV for 9-AGNRs~\cite{talirz2017surface}. We characterized 17-AGNRs using scanning tunneling spectroscopy (STS) (Figure~\ref{fS_Electronic-properties} a-c). The differential conductance (\didv) spectrum shows four main peaks: at -0.9~V (VB-1), -0.2~V(VB), 0.25~V(CB), and 1.3~V (CB+1). Conductance \didv maps of VB and CB show that, similarly to other AGNRs~\cite{ruffieux2012electronic,talirz2017surface}, these are localized along the GNR length with higher intensity at the edges (Figure~\ref{fS_Electronic-properties} d). From these measurements, we extract the experimental bandgap of 17-AGNRs on Au(111) to be approximately 0.45~eV, making it a promising candidate for high-performance devices.

\subsection{Raman characterization of 17-AGNR}
To ensure successful device fabrication, it is crucial to evaluate the quality and robustness of the ribbons during substrate transfer and after device integration, as their structural integrity may be compromised during these steps. Owing to its versatility and high sensitivity to chemical structure, Raman spectroscopy has been extensively employed to characterize $sp^{2}$-hybridized carbon nanomaterials~\cite{Liu2020In,Verzhbitskiy2016Raman,Thapliyal2022concise}. In graphene nanoribbons, it provides insights into GNR width~\cite{Borin2019Surface,Liu2020In,Verzhbitskiy2016Raman,Vandescuren2008Theoretical}, edge structure~\cite{Kim2018Distinguishing,Liu2020In}, length~\cite{Overbeck2019Universal}, and alignment~\cite{Darawish2024Quantifying}. Furthermore, it enables the assessment of large-area uniformity and structural integrity after substrate transfer to dielectric layers and upon device nanofabrication steps. 

The vibrational properties of GNRs give rise to four distinct spectral features: the longitudinal compressive mode (LCM), radial breathing-like mode (RBLM), CH/D modes, and the G mode.\cite{Overbeck2019Universal,Wu2018Raman} The LCM appears at low frequencies (100 cm\textsuperscript{-1}), and its Raman shift is inversely proportional to the ribbon length. The RBLM is another low-frequency mode, and its Raman shift serves as a unique fingerprint of the nanoribbon width. For instance, the RBLMs of 5-, 7-, and 9-AGNRs have been experimentally measured at 531, 399, and 314 cm\textsuperscript{-1}, respectively~\cite{Overbeck2019Optimized}. The CH/D modes, found in the 1100–1500 cm\textsuperscript{-1} range, are characteristic of GNRs' one-dimensionality, with hydrogen-passivated edges breaking the lattice periodicity. Unlike graphene, where these modes emerge due to the presence of defects, they are intrinsic to GNR structures.~\cite{Borin2019Surface,Eckmann2013Raman} The G mode, centered near 1590 cm\textsuperscript{-1} arises from the collective in-plane vibrations of carbon atoms, and it is a signature peak of $sp^{2}$-hybridized carbon systems such as graphene and carbon nanotubes (CNTs). For GNRs, this mode consists of two components: the longitudinal-optical (LO) and the transverse-optical (TO) phonons, reflecting their distinct in-plane vibration directions.~\cite{Overbeck2019Universal,Borin2019Surface,Borin2022Growth}

\begin{figure}[!htbp]
  \centering
  \includegraphics[width=\textwidth]{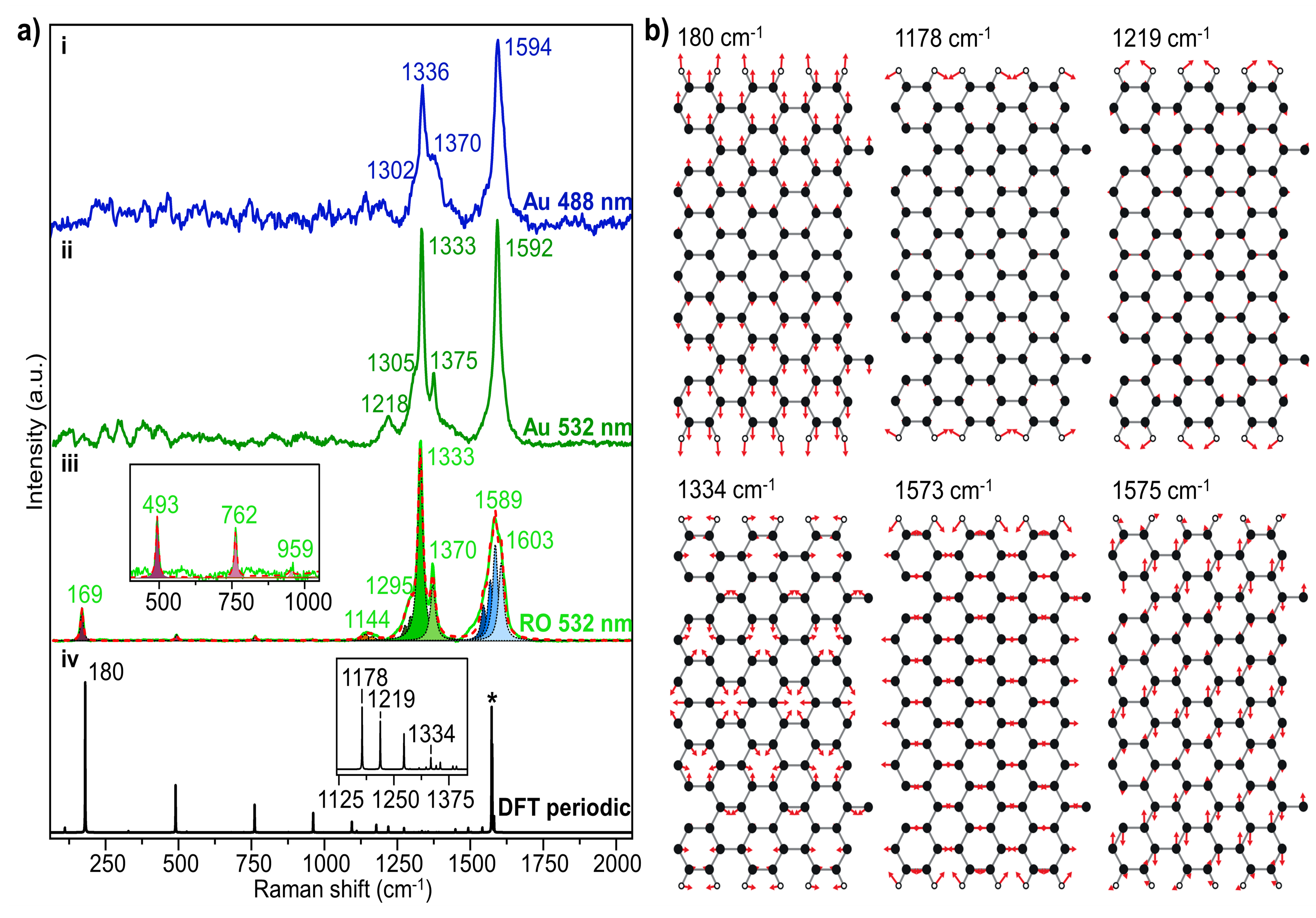}
  \caption{\textbf{Experimental and theoretical Raman spectra of 17-AGNRs}. a) Experimental Raman spectra on Au(111) using 488~nm (i. blue) and 532~nm (ii. green) laser wavelengths and on Raman-optimized (RO) substrate after substrate transfer using 532~nm laser wavelength (iii. light green) with its Lorentzian components, and a simulated Raman spectrum of a periodic 17-AGNR using DFT (iv. black). Four modes are simulated in the G peak: 1541, 1573, 1575, and 1581 cm\textsuperscript{-1}, with the LO mode centered at 1573 cm\textsuperscript{-1} being the most prominent, followed by the TO mode at 1575 cm\textsuperscript{-1} (marked with an asterisk). b) Normal mode analysis for RBLM, representative CH/D modes, and LO and TO phonon modes of the G peak obtained from DFT simulation.}
  \label{f2}
\end{figure}
\FloatBarrier

To investigate the vibrational properties of 17-AGNRs, we measured Raman spectra on Au(111) and upon transfer to a Raman-optimized (RO) substrate (\ce{Al2O3}(40 nm)/\ce{Au}(80 nm)/\ce{SiO2}(285 nm)) using 488, 532, and 785~nm excitation laser wavelengths (\textlambda\textsubscript{ex}) with photon energies of 2.54, 2.33, and 1.58~eV, respectively. (Figures~\ref{f2}a and~\ref{fS_Raman_all}). 

The Raman spectra of 17-AGNRs on Au(111) (Fig. 2a) reveal the characteristic CH/D peaks at $\sim$1305, $\sim$1333, and $\sim$1375 cm\textsuperscript{-1} as well as a G peak at $\sim$1592 cm\textsuperscript{-1}. Notably, a peak centered at 1218 cm\textsuperscript{-1} is observed upon \textlambda\textsubscript{ex} = 532~nm (2.33~eV) excitation but is absent at \textlambda\textsubscript{ex} = 488~nm (2.54~eV), indicating selective resonance with the 2.33~eV excitation.

To gain a deeper understanding of the origin of these vibrational modes, we performed periodic DFT calculations on an infinitely long 17-AGNR (Figure~\ref{f2}a panel iv). While the calculated intensities do not perfectly match experimental values, likely due to the absence of resonance effects and an accurate electron-phonon coupling description, the peak frequencies are in good agreement. Normal mode analysis (Figures~\ref{f2}b and~\ref{fS_modes}) confirms that the peaks at 1305–1375 cm\textsuperscript{-1} arise from a mixture of carbon-hydrogen vibrations and ring breathing within the 1D GNR lattice. Additionally, it shows that the peak centered at 1218 cm\textsuperscript{-1} is associated with the carbon-hydrogen bending mode at the edges.

Low-frequency modes, such as the LCM and RBLM, can be suppressed on gold substrates due to their weak signal intensity and the substrate's influence on the specific collective motions characteristic of these modes. As demonstrated for other AGNRs~\cite{Overbeck2019Optimized}, transferring ribbons to an RO substrate enhances these modes by decoupling the ribbons from the metallic surface. Upon polymer-free transfer of 17-AGNRs (see Methods), we observe distinct low-frequency modes at 169, 493, 762, and 959 cm\textsuperscript{-1} with \textlambda\textsubscript{ex}=532~nm. These modes are associated with the RBLM and its overtones. When measured at \textlambda\textsubscript{ex}=488~nm, additional modes appear at the frequencies below 150 cm\textsuperscript{-1}, which will be discussed in the next section.

The presence of the RBLM and negligible changes in the overall peak positions before and after the transfer indicate that 17-AGNRs are stable in ambient conditions and their structure is preserved during the polymer-free transfer procedure.~\cite{Borin2019Surface} Guided by our DFT and normal mode analyses (Figures~\ref{f2}b and~\ref{fS_modes}), we deconvoluted the experimental Raman spectra into their Lorentzian components (Figures~\ref{f2}a panel iii,~\ref{fS_Raman_all} panels i, ii, iv and v, and table~\ref{tS2}). The positions of the simulated spectra and the deconvoluted components agree relatively well, capturing the low-frequency and high-frequency modes. 

Interestingly, the number and the widths of the constituent modes vary before and after the transfer (Figure~\ref{f2}a, Table~\ref{tS2}). Focusing on the deconvoluted peaks from the spectra measured on Au(111) and on RO at \textlambda\textsubscript{ex}=532~nm (table \ref{tS2}), CH/D modes centered at 1219, 1410, and 1437 cm\textsuperscript{-1} become less prominent, whereas the one at 1277 cm\textsuperscript{-1} newly appears after the transfer of the ribbons from gold to the RO substrate. 
In the G mode region, the splitting of the G peak into the different components is observed (table \ref{tS2}). This is captured by both the deconvolution (1545, 1568, 1586, and 1606 cm\textsuperscript{-1}, on the RO substrate with \textlambda\textsubscript{ex}=532~nm) and normal mode analysis (1541, 1573, 1575, and 1581 cm\textsuperscript{-1}). Note that the intensities of CH/D modes are underestimated in the calculated Raman spectrum, likely due to the absence of resonance effects and precise electron-phonon coupling, which are not taken into account in this simulation.
The modes at 1545, 1568, and 1606 cm\textsuperscript{-1} appear after the transfer in addition to the one at 1593 cm\textsuperscript{-1} on gold (1586 cm\textsuperscript{-1} on RO substrate). In addition, the width of the G mode centered at 1593 cm\textsuperscript{-1} decreases by 4 cm\textsuperscript{-1}, and those of the CH/D modes at 1305, 1333, and 1375 cm\textsuperscript{-1} increase by 3 cm\textsuperscript{-1} for the G modes. These may originate from multiple sources, such as GNRs interacting differently with different substrates (substrate effect) or variations in the length of the ribbons, as shown previously in 5-AGNRs where the Raman shift of the LO branch changes with GNR length, particularly for shorter ribbons~\cite{Borin2022Growth}. In addition, discrepancies in the vibrational modes resolved at different excitation energies could be due to different modes getting excited differently depending on the excitation energy. Similar behaviors have already been observed for 5-, 7-, and 9-AGNRs.\cite{Borin2019Surface,Borin2022Growth} 

\subsection{Length-dependent Raman studies}

\begin{figure}[!htbp]
  \centering
  \includegraphics[width=\textwidth]{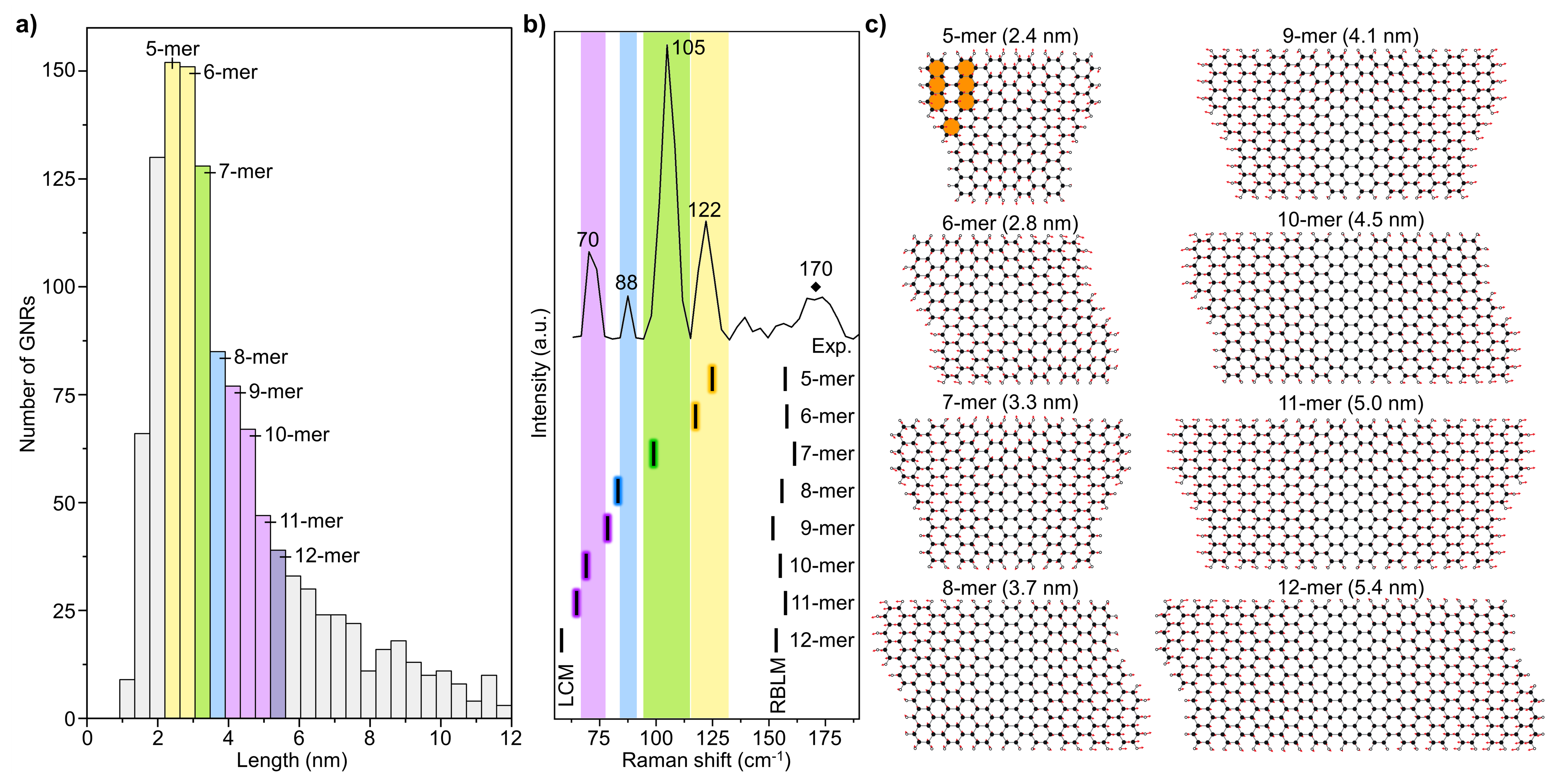}
  \caption{\textbf{Low frequency mode analysis.} a) Length distribution of the 17-AGNR sample measured by Raman. The histogram is truncated at 12~nm (bin size: 0.4~nm). 5- to 12-mers are highlighted in colors that represent the ranges of the LCM positions. b) Low-frequency region of the Raman spectrum of 17-GNRs on a 30~nm-thick \ce{Al2O3} dielectric layer measured at \textlambda\textsubscript{ex} = 488~nm (Exp., black). The RBLM from the experimental spectrum is marked with a black diamond ($\Diamondblack$). Below, the theoretical positions of the LCMs and RBLM of 5- to 12-mer 17-AGNRs are marked with vertical lines. c) Atomic displacement profiles of LCMs for 5- to 12-mer 17-AGNRs modeled with REBOII. A monomer unit is highlighted in orange in a 5-mer 17-AGNR, and in the parentheses are the simulated sizes.}
  \label{f3}
\end{figure}

As observed in our scanning tunneling microscopy study of samples with varying initial precursor coverages, the final GNR length can differ significantly. To investigate this length variation using Raman spectroscopy~\cite{Overbeck2019Universal,Borin2023On} we synthesized a 17-AGNR sample with an average length of 4.5~nm (Figure~\ref{f3}a), comparable to the sample shown in Figure 1b. Raman spectra of this sample, measured with an excitation wavelength of \textlambda\textsubscript{ex} = 488~nm on a 20~nm-thick \ce{Al2O3} gate dielectric, reveal multiple peaks below 150 cm\textsuperscript{-1} (figure~\ref{f3}b). 

We observe four low-frequency peaks at 70 (violet), 88 (blue), 105 (green), and 122 (yellow) cm\textsuperscript{-1} (Figure~\ref{f3}b). To identify the origin of these modes and verify their correlation with GNR length, we employed the second-generation reactive empirical bond-order (REBOII) potential~\cite{Brenner_1990,Brenner_2002} to simulate the Raman spectra of 17-AGNRs with different lengths, ranging from 5 to 12 monomers (5- to 12-mer 17-AGNRs, table~\ref{tS3}). Unlike density functional theory (DFT), which solves the Schr\"odinger equation through self-consistent iterations to model electronic interactions, REBOII employs predetermined mathematical functions to describe atomic interactions. This makes it a computationally efficient alternative that balances accuracy and resource demands for simulating vibrational modes in finite systems (see Methods). To evaluate the reliability of the REBOII model, we compared DFT- and REBOII-derived Raman spectra of 7- and 8-units long 17-AGNRs (7- and 8-mer 17-AGNRs). Figure~\ref{fS_DFT-vs-REBO} shows that while significant shifts occur in the high-frequency modes, the low-frequency modes exhibit negligible variation between the two methods. This confirms that the REBOII potential accurately captures the interatomic interactions between carbon atoms near equilibrium. 

Our simulations indicate that the LCMs of 9-, 10-, and 11-units long 17-AGNRs (9-, 10-, and 11-mer 17-AGNRs, corresponding to lengths between 4 and 5~nm) contribute to the peak observed at 70 cm\textsuperscript{-1}. The histogram in Figure~\ref{f3}a shows a comparable amount of ribbons consisting of 9, 10, and 11 monomers, suggesting that their combination likely contributes to the rise of this peak. The peak at 88 cm\textsuperscript{-1} is assigned to the LCM of the 8-mer (3.7~nm) 17-AGNR, while the peaks at 105 cm\textsuperscript{-1} and 122 cm\textsuperscript{-1} are attributed to the LCMs of the 7-mer (3.3~nm) and 5- and 6-mers (2.4-2.8~nm) 17-AGNRs, respectively. The strong agreement between the Raman experimental data and simulation reflects the length distribution observed in the STM-derived histogram, particularly in the sub-7~nm range where Raman spectroscopy is most sensitive to changes in ribbon length~\cite{Overbeck2019Universal}.

\FloatBarrier

Another notable feature is the broad RBLM peak (marked with a black diamond) observed in figure~\ref{f3}b. We attribute this broadening to two factors: the variation in RBLM positions for 5- to 12-mer ribbons, which range from  154-162 cm\textsuperscript{-1} and the splitting of the RBLM into normal modes with diagonal atomic displacement in short GNRs\cite{Overbeck2019Universal} (Figure~\ref{fS_rebo}).

Finally, it is important to note that the LCMs corresponding to ribbons of length 5.4~nm or longer (ribbons with 12 or more monomer units) are not visible in this spectrum despite accounting for approximately 26.5\% of the sample. This absence is likely due to the proximity of their vibrational frequencies to the Rayleigh scattering line, which limits their detectability in the Raman measurement.

\subsection{Device integration and transport measurements}
\begin{figure}[!htbp]
  \centering
  \includegraphics[width=\textwidth]{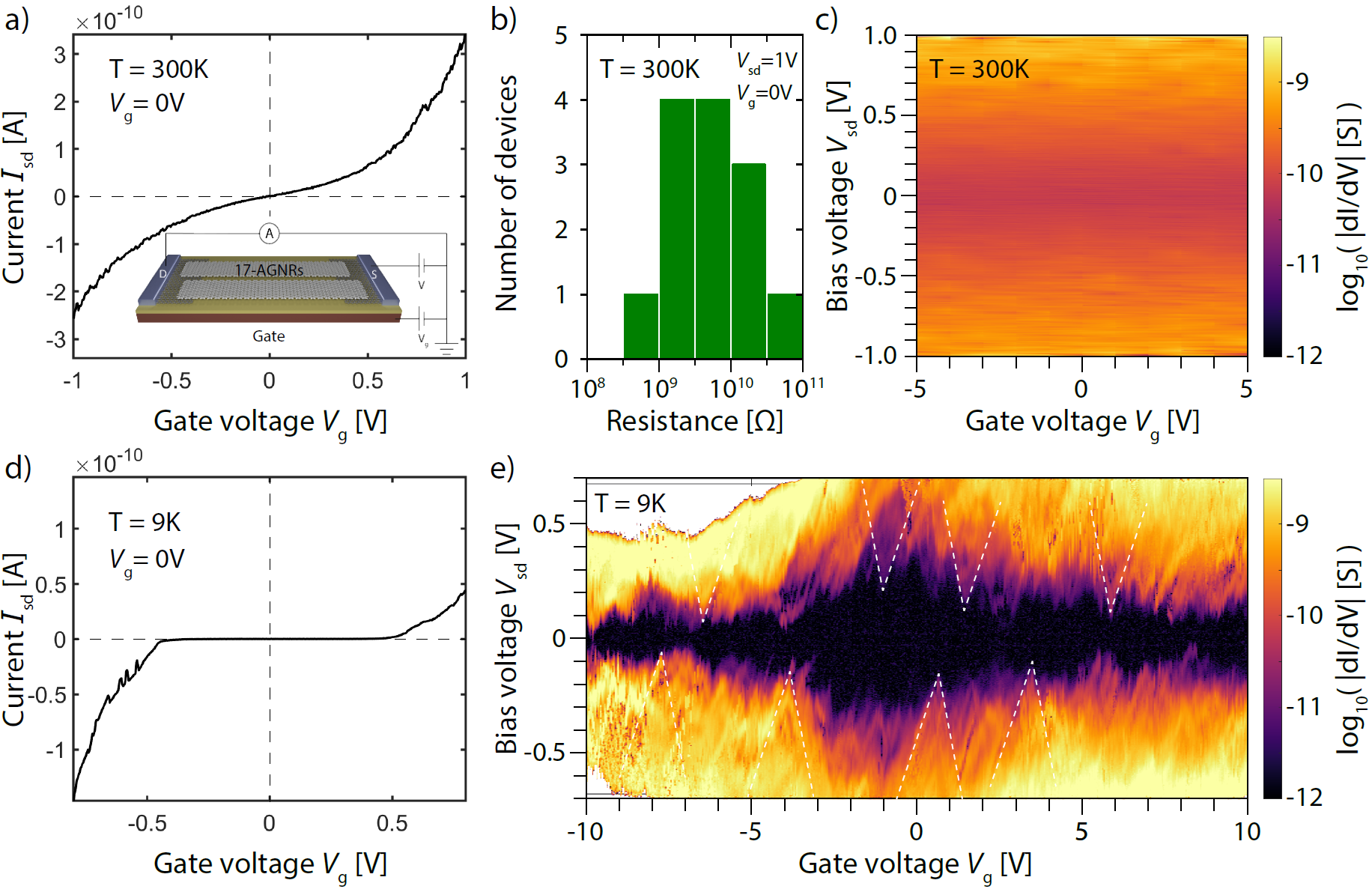}
  \caption {\textbf{Electrical characterization of 17-AGNR devices.} a) Current-voltage ($I-V$) trace of a typical device at room temperature. The inset shows the schematic of the device and measurement layout. b) Histogram of the room temperature device resistance for all 17-AGNR devices extracted at a bias voltage of 1 V due to the non-linearity of the $I$-$V$ curve. c) Differential conductance \didv map of the device in (a) at room temperature. d) $I-V$ trace of the same device as in (a) recorded at 9 K. e) \didv map as a function of applied bias and gate voltages. The black region indicates the Coulomb blockade. White dashed lines are guides to the eye.}
  \label{f4}
\end{figure}

After demonstrating that 17-AGNRs can grow long enough to bridge source and drain and withstand the wet-transfer process, we integrated the GNRs into devices. The selected device structure is based on bottom electrodes made of graphene and defined using electron-beam lithography (EBL) and reactive ion etching (RIE), yielding nanogaps in the range of 15-20~nm~\cite{Braun2021Optimized}. The devices were fabricated as follows (more details in Methods):
1. Silicon/silicon dioxide (\ce{Si/SiO2}) with an oxide thickness of 285~nm served as the basis for the device fabrication. 2. A local gate is defined on top using standard EBL, covered using a $\sim$20~nm thick \ce{Al2O3} gate dielectric. (Figure~\ref{fS_device-assessment} and \ref{f4}a inset) 3. Platinum lines on top serve as contacts to graphene electrodes. 4. Graphene electrodes are chosen due to their flatness and their proven capability to contact molecules~\cite{El2019Robust} as well as graphene nanoribbons.~\cite{Braun2021Optimized,El2020Controlled,Niu2023Exceptionally,chen2024porphyrin}

\FloatBarrier

Room-temperature charge transport measurements reveal $I-V$ characteristics with source-drain currents reaching the nanoamperes range at an applied bias voltage of 1 V (Figure~\ref{f4}a). The curves are S-shaped, suggesting a Schottky-like barrier is present at the GNR/electrode interface. A thermal annealing step has been proposed by Braun et al. to improve the GNR/electrode interface~\cite{Braun2021Optimized}. For 9-AGNRs, this step leads to improved device performance, primarily due to local rearrangements and water removal at the GNR/graphene interface. However, such a step was not performed as a degradation of the 17-AGNRs was observed using Raman spectroscopy after annealing (Figure~\ref{fS_Raman-on-device}).

Figure~\ref{f4}b shows the histogram of the conductance of 13 different 17-AGNR devices measured, with a peak around 1-10 GOhm. The resistance values were extracted at an applied bias voltage of 1 V due to the nonlinearity of the $I-V$ characteristics. The spread in resistance values is attributed to the device-to-device variations in GNR film quality (orientation of GNRs or coverage) or minor variations in electrodes (nanogap size, folds, wrinkles, or contaminants).

$I-V$s were recorded as a function of gate voltage, but little to no effect was observed (Figure~\ref{f4}c) for a representative device. This lack of gate dependence at room temperature is consistent with previous reports for other low-bandgap GNRs, for which no appreciable gate-dependent transport has been observed.~\cite{El2020Controlled,Sun2020Massive}

At low temperature, the $I-V$ characteristics exhibit pronounced non-linearity with distinct step-like features (Figure~\ref{f4}d). These steps manifest as resonances in the differential conductance (\didv), indicating transport through quantized energy levels. Figure~\ref{f4}e shows a color map of the differential conductance as a function of bias and gate voltage. Regions of high and low conductance are clearly visible, reminiscent of Coulomb blockade and single-electron tunneling regimes, respectively. However, despite the presence of sharp resonant features (white dashed lines), no clear Coulomb diamonds with well-defined crossing points are observed. This absence suggests that electronic transport occurs simultaneously through multiple conduction channels, implying that several 17-AGNRs contribute to the measured current. Given the graphene electrode geometry having widths of approximately 400~nm, it is likely that multiple 17-AGNRs bridge the electrodes both in parallel and in series. Additionally, localized states in the electrodes or charge traps within the oxide may further influence transport characteristics. Although such complexity prevents clear Coulomb diamond formation, these observations confirm that gate-tunable resonant transport can indeed occur at low temperatures, in contrast to room-temperature measurements.

\section{Conclusion}
In this study, we optimized the on-surface synthesis conditions to produce long and high-quality 17-AGNRs by on-surface synthesis approach. We identified two key factors critical to achieving this: a gradual annealing process and a template-like effect induced by monomer clustering at high precursor coverage. Raman spectroscopy, performed both on the Au(111) growth substrate and after transfer to oxide-based device platforms, revealed that 17-AGNRs exhibit remarkable structural robustness despite their low bandgap. Furthermore, we demonstrate that Raman spectroscopy can be used to efficiently characterize GNR length by identifying several LCM modes. Owing to their optimized length and robustness, 17-AGNRs were successfully integrated into electronic devices. We investigated the transport properties of these GNRs in a FET geometry, where gate-independent transport was observed at room temperature, and sharp transport features were observed at cryogenic temperatures. Our findings highlight that 17-AGNRs, the widest AGNRs experimentally realized within the $3p + 2$ family, can be synthesized with significantly increased lengths and quality. This, combined with their structural robustness, allows the investigation of their electronic properties in an FET configuration.

\section{Methods}

\textbf{\textit{Precursor synthesis}}
The synthesis of \textbf{BADBB} was performed in solution through the Suzuki coupling of 1,4-dibromo-2,3-diiodobenzene and anthracen-2-ylboronic acid as previously described~\cite{Yamaguchi2020Small,dumslaff_2017}. The final product quality was confirmed by Nuclear Magnetic Resonance (NMR) analysis.

\textbf{\textit{Sample preparation}}
The Au(111) single crystals (MaTeck GmbH) were prepared by iterative Ar\textsuperscript{+} sputtering and annealing cycles. Prior to the sublimation of other species, the surface structure and cleanliness were checked by STM imaging. The molecular precursor \textbf{BADBB} was loaded into quartz crucibles of a home-built evaporator and, after proper degassing, sublimated at approximately 250~°C onto a room-temperature substrate. After deposition, the precursors were annealed by applying a constant heating ramp of roughly 2~°C/min until 400~°C to favor the formation of long 17-AGNRs.

\textbf{\textit{STM characterization}}
STM measurements were performed with a commercial variable-temperature STM from Scienta Omicron operated at room temperature and base pressure below 5×10\textsuperscript{-11} mbar. In addition, due to their high mobility, measurements of \textbf{BADBB} precursors were acquired at LN\textsubscript{2} temperature (77 K), and spectroscopy data at LHe temperature (4.2 K). STM images were acquired in constant-current mode (overview and high-resolution imaging), \didv spectra were acquired in constant-height mode, and the feedback-loop off parameters are reported in the captions. Differential conductance \didv spectra were obtained with a lock-in amplifier. The data were processed with Wavemetrics Igor Pro software.

\textbf{\textit{DFT-AiiDA}} 
DFT calculations were executed using the AiiDAlab applications~\cite{yakutovich2021aiidalab}, based on AiiDA workchains~\cite{pizzi2016aiida} designed for the DFT code CP2K~\cite{hutter2014cp2k} (systems adsorbed on gold) and for the DFT code Quantum Espresso (bandstructure calculations)~\cite{giannozzi2009quantum}. Surface-adsorbate setups were modeled within a periodic slab scheme. The simulation cell included four Au atomic planes along the [111] orientation. Hydrogen atoms passivated one face of the slab to mitigate Au(111) surface states. A 40 \AA  vacuum layer was included to isolate the system from its periodic images along the axis orthogonal to the slab. Electronic wavefunctions were represented via TZV2P Gaussians basis sets for C and H, and DZVP for Au. Plane-waves basis set cutoff for the charge density was set at 600 Ry. Norm-conserving Goedecker–Teter–Hutter pseudopotentials were employed. The Perdew-Burke-Ernzerhof (PBE) generalized gradient approximation (GGA)~\cite{perdew1996generalized} approximation for the exchange correlation functional was used, and Grimme’s D3~\cite{grimme2010consistent} van der Waals corrections were included. Au supercells varied in size depending on the adsorbates, ranging from 20.64 × 28.12 \AA \textsuperscript{2} (corresponding to 384 Au atoms) to 35.37 × 43.42 \AA \textsuperscript{2} (936 Au atoms). Geometry optimizations were performed with the bottom two atomic planes constrained while relaxing others until forces were below 0.005~eV \AA.

For the bandstructure calculations, ultrasoft pseudopotentials from the Standard solid-state pseudopotentials (SSSP)~\cite{lejaeghere2016reproducibility} were employed to model the ionic potentials. A cutoff of 50 Ry (400 Ry) was used for the plane wave expansion of the wave functions (charge density). The simulation cell contained 15 \AA of vacuum in the non-periodic directions to minimize interactions among periodic replicas of the system. The thickness of the vacuum region, the sampling of the Brillouin zone (BZ), and the cutoff ensure the convergence of the computed band structures. The atomic positions of the ribbon atoms and the cell dimension along the ribbon axis were optimized until forces were lower than 0.002~eV/\AA and the pressure in the cell was negligible. The band structures are aligned to the vacuum level computed from the average electrostatic potential in the vacuum region.

\textbf{\textit{Substrate transfer}}
The GNR/Au(111)/mica samples were placed on an HCl (\textit{aq}). Within 15 minutes, the GNR/Au film cleaved from the mica substrate, after which the solution was diluted with water. The floating GNR/Au(111) film was then immersed in the diluted solution with the target substrate until it was fully submerged. The adhesion between the Au film and the target substrate was enhanced by applying a drop of ethanol to the Au film, followed by annealing at 100 °C for 10 minutes on a hot plate. The Au film was etched away using potassium iodide-based gold etchant (Sigma-Aldrich CAS No. 7681-11-0), resulting in GNRs on a target substrate.~\cite{Borin2019Surface}

\textbf{\textit{Raman spectroscopy}}
After the sample was prepared and characterized using STM, it was transferred into a vacuum apparatus with optical access to the Raman microscope. Raman spectra were measured in back-scattering geometry using a Witec Alpha 300 R confocal Raman microscope, with 488, 532, and 785~nm excitations at 600 g/mm (488 and 532~nm) and 300 g/mm (785~nm) grating. For the respective excitations, the spectra were measured at 10, 5, and 2 mW with integration times of 10, 10, and 15 seconds, respectively, for the sample on a gold surface. The samples on RO were measured at 2, 5, and 10 mW, with integration times of 5, 1, and 5 seconds, respectively. The sample on a \ce{Al2O3} gate dielectric of a device was measured with 488~nm excitation at 1 mW for 10 seconds. The Raman spectra were recorded at 100 points and averaged. All spectra were collected using a 50x LD objective (Zeiss, NA = 0.55). All Raman spectra were collected at the vacuum level ranging from 1-30 × 10\textsuperscript{-6} mbar. The data were processed with OriginPro.

\textbf{\textit{Raman simulation}}
The phonon calculation needed for Raman was carried out by Phonopy~\cite{Togo_2015}; an open-source package to calculate phonons within the harmonic and quasi-harmonic approximation. Phonopy was paired with a calculator for the computation of the atomic forces obtained within a finite displacement scheme. For Raman calculation with REBO forces, the REBOII potential~\cite{Brenner_2002} was used for the calculation of the forces. REBOII potential is a classical hydrocarbon potential energy expression, which allows for bond making and breaking with appropriate changes in atomic hybridization~\cite{Brenner_1990, Brenner_2002}. Additionally, the bond polarizability model~\cite{Guha_1996, Saito_2010} was used for the calculation of the polarizability derivatives to obtain the Raman intensities.
For Raman calculation of the periodic structure with DFT forces, structure relaxation was first performed by GPAW~\cite{Mortensen_2005, Enkovaara_2010}, which is a DFT Python code based on the projector-augmented wave (PAW) method and the atomic simulation environment (ASE)~\cite{Larsen_2017}. For relaxation, a plane-wave basis set was used with a plane-wave energy cutoff of 450~eV with a $\Gamma$-centered 9 × 1 × 1 Monkhorst-Pack Brillouin-zone sampling. Then, the phonon calculation was carried out by Phonopy using a 3 × 1 × 1 supercell with a displacement distance of 0.02 \AA. The calculation of the force constants was performed by GPAW using the same plane-wave energy cutoff as used in the relaxation with a $\Gamma$-centered 4 × 1 × 1 Monkhorst-Pack sampling. The exchange-correlation functional used in these calculations was the GGA in the PBE formulation~\cite{Perdew_1996}.

\textbf{\textit{Device fabrication}}
Highly p-doped \ce{Si/SiO2} (285~nm) serves as a base substrate for the device fabrication and serves as a global back gate (set to V\textsubscript{gb} = 0~V). A thin gate electrode (Ti/Pd, 1/6~nm) is placed in the central region of the device and is covered by an atomic layer deposition (ALD)-grown dielectric layer (\ce{Al2O3}, 20~nm). This allows for electrostatic gating of the 17-AGNRs. Palladium is chosen for the gate because it forms uniform, thin layers with low surface roughness.~\cite{nazarpour2010material} Metal wires (Ti/Pt, 5/35~nm) serve as contact to the graphene electrodes. Platinum was chosen since it allows the transfer of GNRs grown on Au(111)/mica substrates, where the transfer process involves a gold etching step that does not attack platinum. In a last step, chemical vapor deposition (CVD)-grown graphene is placed on top and patterned using EBL.
The cleanliness of the devices is confirmed using atomic force microscopy, and electrical separation is performed using electrical transport characterization, as described below. 

\textbf{\textit{Transport}}
All electronic measurements are performed under vacuum conditions (\textless 10\textsuperscript{-6} mbar) in a commercially available probe station (Model CRX-6.5K, Lake Shore Cryogenics). A data acquisition board (ADwin-Gold II, J\"ager Computergesteuerte Messtechnik GmbH) is employed to apply and read the voltages from the $I-V$ converters (DDPCA-300, FEMTO Messtechnik GmbH and DELFT ELECTRONICS).

\section*{Author contribution}
G.B.B. conceived and supervised the project. J.H.H., N.B., and G.B.B. performed the growth and characterization of GNRs. J.H.H. and N.B. analyzed the data. M.F. conducted Raman simulations under the supervision of V.M. O.B., M.S., and R.F., fabricated devices, and carried out transport measurements under the supervision of M.L.P. T.D. and H.H. synthesized the molecular precursors under the supervision of H.Y. and A.N. N.B. and C.P. performed the DFT calculations. J.H.H., N.B., and G.B.B. wrote the manuscript with input from all co-authors. G.B.B., P.R., R.F., and M.C. acquired funding and revised the manuscript. 

\section*{Statement of conflict of interest}
The authors declare no competing financial interest.

\section*{Acknowledgements}

J.H.H., N.B., G.B.B., P.R., and R.F. greatly appreciate the financial support from the Werner Siemens Foundation (CarboQuant).
O.B. and M.C. acknowledge funding by the EC H2020 FET Open project no. 767187 (QuIET).
M.C. acknowledges funding by the Swiss National Science Foundation under the Sinergia grant no. 189924 (Hydronics).
M.F. and V.M. acknowledge the support of the Focus Center at Rensselaer Polytechnic Institute, a NYS Department of Economic Development award (No. A21-0125-002 and RPI designation A50634.2326).
M.L.P. acknowledges funding by the Swiss National Science Foundation (SNSF) under the Spark project no. 196795 and the Eccellenza Professorial Fellowship no. PCEFP2\textunderscore203663. M.L.P. is also grateful for funding from the Swiss State Secretariat for Education, Research and Innovation (SERI) under contract number MB22.00076.
H.H. and H.Y. are also grateful for JSPS KAKENHI grant no.  JP20H02816 (H.H.), JP24K01576 (H.H.), JP20H05833 (H.Y.), and JP25K01751 (H.Y.).
G.B.B., P.R., and R.F. acknowledge funding by the Swiss National Science Foundation under grant no. 212875 (SYNC) and the European Union Horizon 2020 research and innovation program under grant agreement no. 881603 (GrapheneFlagship Core 3). G.B.B. also greatly appreciates the European Union’s Horizon Europe research and innovation program under grant agreement no. 101099098 (ATYPIQUAL) and the Swiss National Science Foundation grant no. 200021E-219172/1 (GRAAL).
The authors are grateful for access to the Scanning Probe Microscopy User Lab at Empa for the AFM measurements, as well as the support provided by the staff of the Binnig and Rohrer Nanotechnology Center (BRNC) for device fabrication.

\bibliography{references.bib}

\clearpage
\section{Supporting Information}
\setcounter{figure}{0}
\setcounter{table}{0}
\renewcommand{\thefigure}{S\arabic{figure}}
\renewcommand{\thetable}{S\arabic{table}}

\begin{figure}[!htbp]
    \centering
    \includegraphics[width=\textwidth]{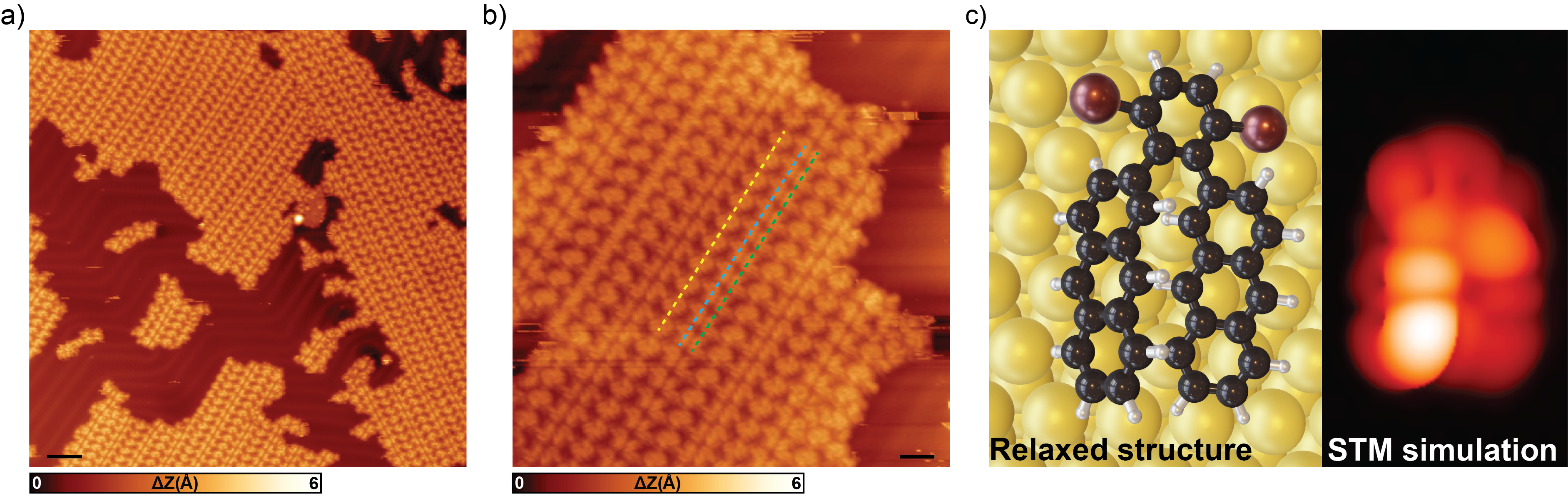}
    \caption{\textbf{BADBB on Au(111)}. a) Large-scale STM topography of \textbf{BADBB} deposited on Au(111). The molecules self-assemble into large islands (scale bar: 10~nm, scanning parameters: -1 V and 50~pA). b) Zoomed-in STM image on an island. The molecules have three main orientations (scale bar: 5~nm, scanning parameters: -1 V and 50~pA). c) DFT optimized structure (left) and corresponding STM simulation (right) of \textbf{BADBB} on Au(111). Due to steric hindrance between the anthracene units, one of them protrudes away from the surface. All the STM images have been acquired at 77~K.}
    \label{fS_BADBB-on-Au}
\end{figure}

\begin{figure}[!htbp]
\centering
    \includegraphics[width=\textwidth]{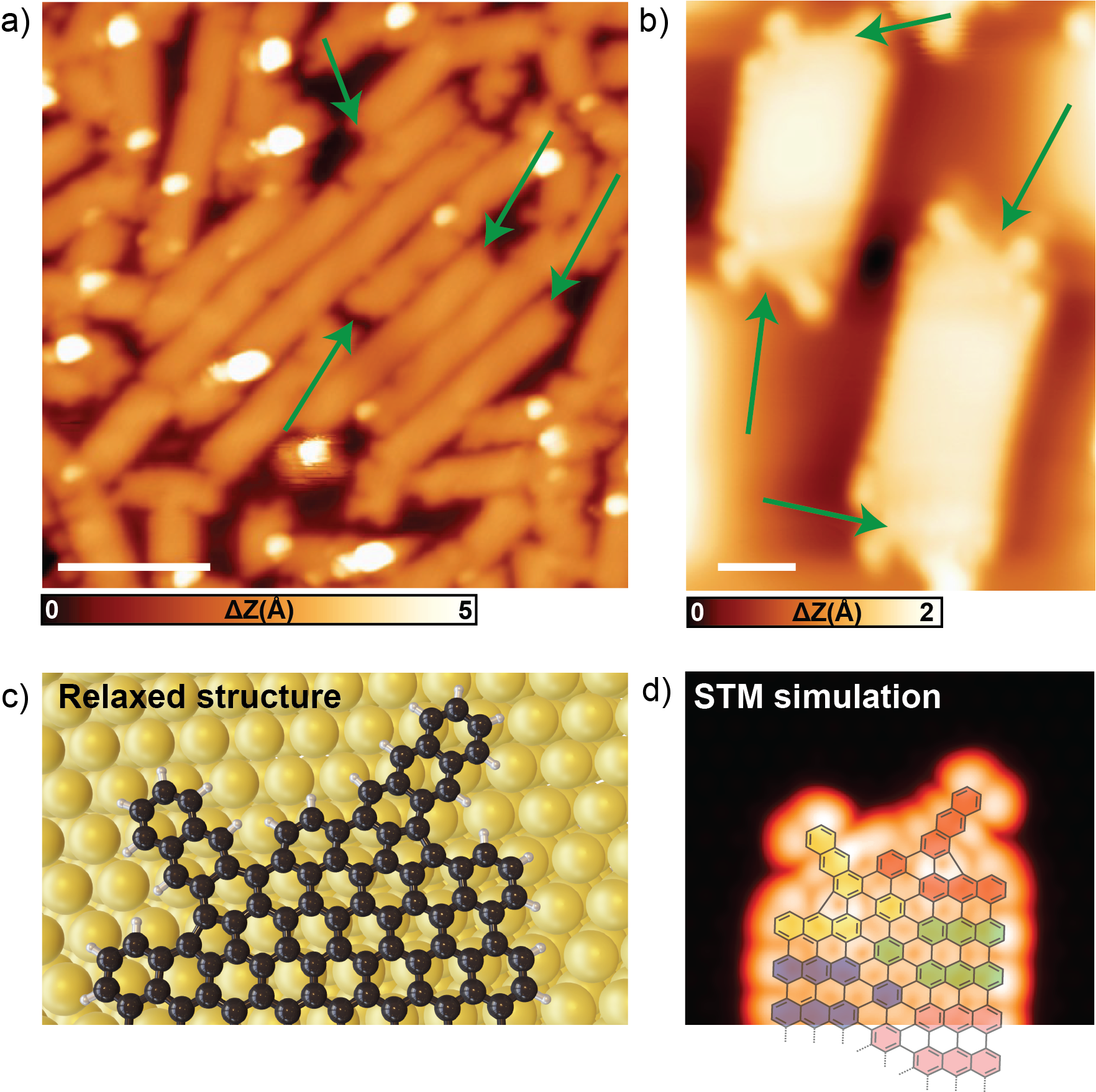}
    \caption{\textbf{End segments of 17-AGNR}. a) STM topography of high surface coverage with long 17-AGNRs on Au(111). The green arrows point to the ends of the nanoribbons which are characterized by two protrusions rather than the expected staggered geometry. (scale bar: 5~nm, scanning parameters: -1~V and 50~pA). STM image measured at room temperature. b) STM topography of short 17-AGNRs on Au(111). The ribbon termini are always characterized by a specific double structure (scale bar: 1~nm, scanning parameters: -0.1~V and 50~pA). STM images acquired at 4.3~K.  c) DFT relaxed structure (left) and (d) corresponding STM simulation of 17-AGNR ends. The specific double feature structure of the ribbon termini is explained by a "flipped" conformation of the two terminal precursor units, resulting in the flipped anthracene units protruding out of the ribbon structure after the thermally induced cyclization step.}
    \label{fS_End-segment}
\end{figure}

\begin{figure}[!htbp]
\centering
    \includegraphics[width=\textwidth]{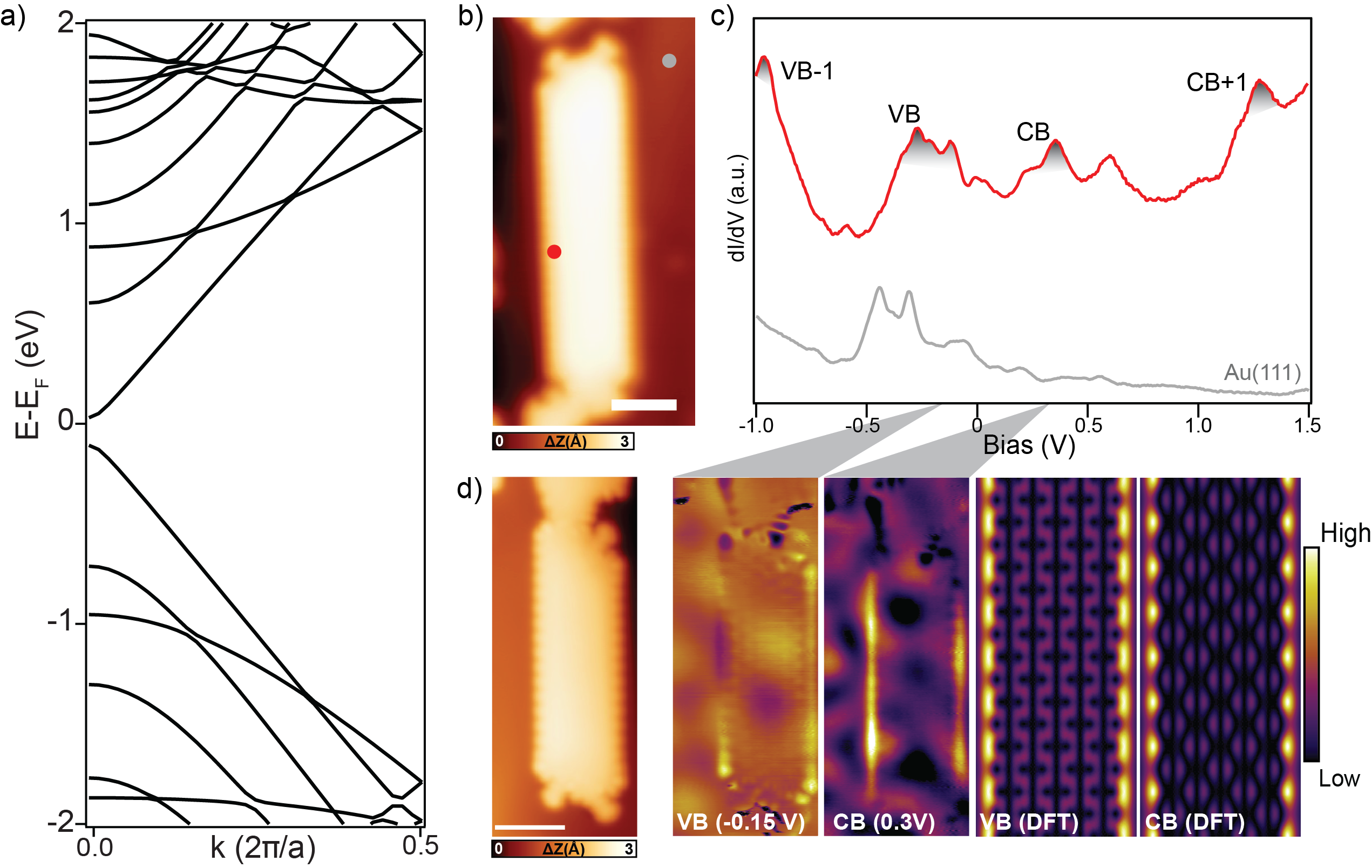}
    \caption{\textbf{Electronic properties of 17-AGNR}. a) Gas-phase DFT calculation for an infinitely long 17-AGNR segment. The predicted bandgap is 0.14~eV.
    b) STM topography of a 17-AGNRs segment on Au(111) (scale bar: 2~nm, scanning parameters: -1~V and 50~pA). c) \didv spectra taken at the edges (panel b). Four main peaks are visible, corresponding to VB-1 (-0.95~V), VB (-0.15~V), CB (0.3~V), and CB+1 (1.28~V). (Open feedback parameters: -1.0~V, 500~pA; Vrms: 20~mV. In gray, spectra on Au(111) for comparison). d) STM topography and corresponding constant current conductance maps of VB (-0.15~V) and CB (+0.3~V)  (DFT simulations on long 17 AGNRs unit are also reported). Both simulations and experimental maps shows that these states spread over the ribbon length. The STM images and spectroscopy curves have been performed at 4.3 K.}
    \label{fS_Electronic-properties}
\end{figure}

\begin{figure}[!htbp]
    \centering
    \includegraphics[width=\textwidth]{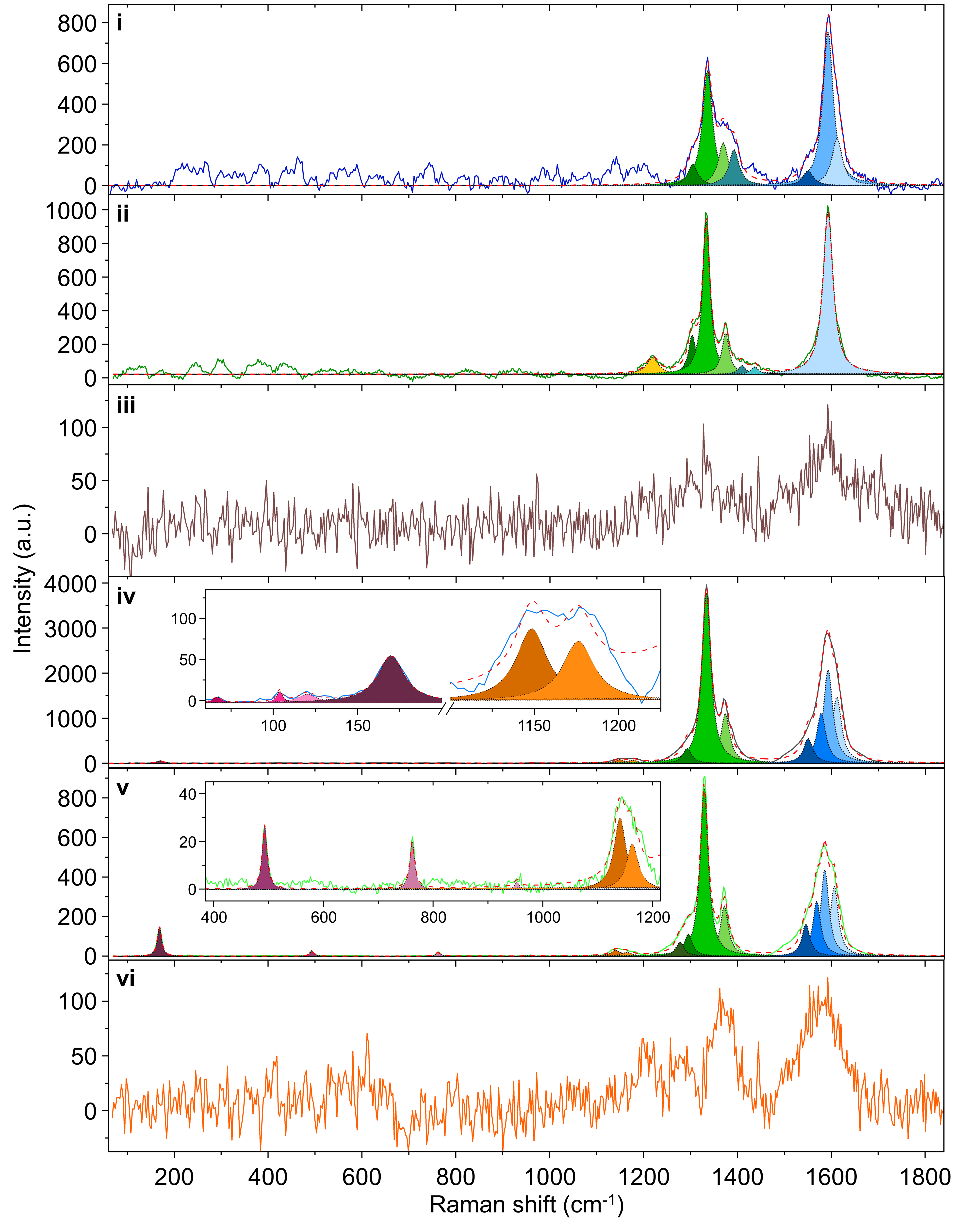}
    \caption{\textbf{Experimental Raman spectra of the 17-AGNRs} on Au(111) collected with i. 488~nm, ii. 532~nm, and iii. 785~nm laser excitation and on RO substrate with iv. 488~nm, v. 532~nm, and vi. 785~nm lasers. Panels i, ii, iv, and v include deconvoluted Lorentzian constituents. Insets in panels iv and v show the regions below $\sim$1225 cm\textsuperscript{-1}. See tables~\ref{tS1} and~\ref{tS2} for the peak information. Note that there are shoulders next to the strongest G mode at $\sim$1593 cm\textsuperscript{-1} in panel ii, which do not deconvolute nicely. This indicates that the other modes are present but not intensely excited as they are on the RO substrate in panel v.}
    \label{fS_Raman_all}
\end{figure}

\begin{figure}[!htbp]
    \centering
    \includegraphics[width=\textwidth]{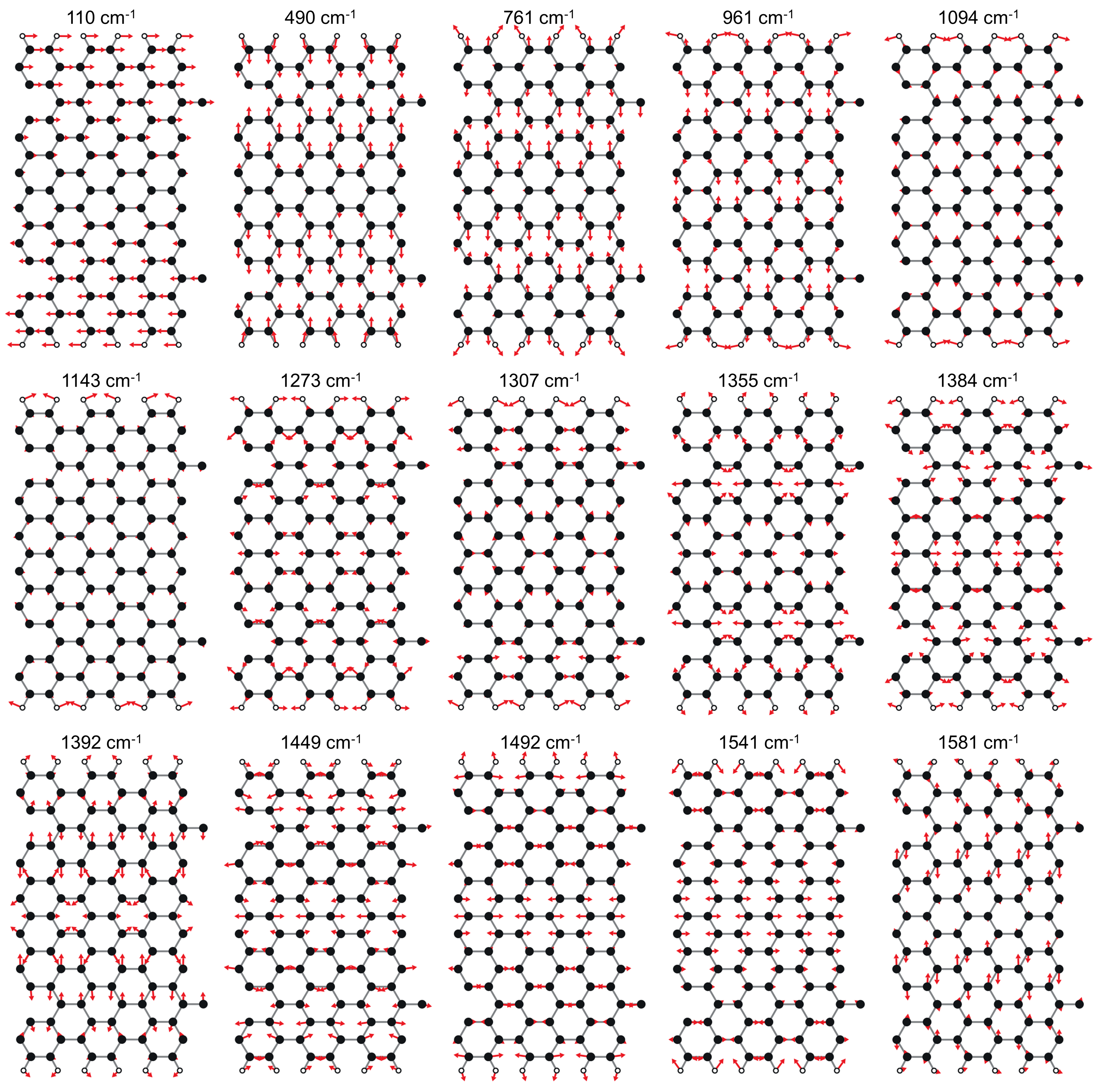}
    \caption{\textbf{Raman normal mode analysis of a periodic structure from the DFT calculation.} Red arrows indicate amplitudes and directions of atomic displacements.}
    \label{fS_modes}
\end{figure}

\begin{figure}[!htbp]
    \centering
    \includegraphics[width=\textwidth]{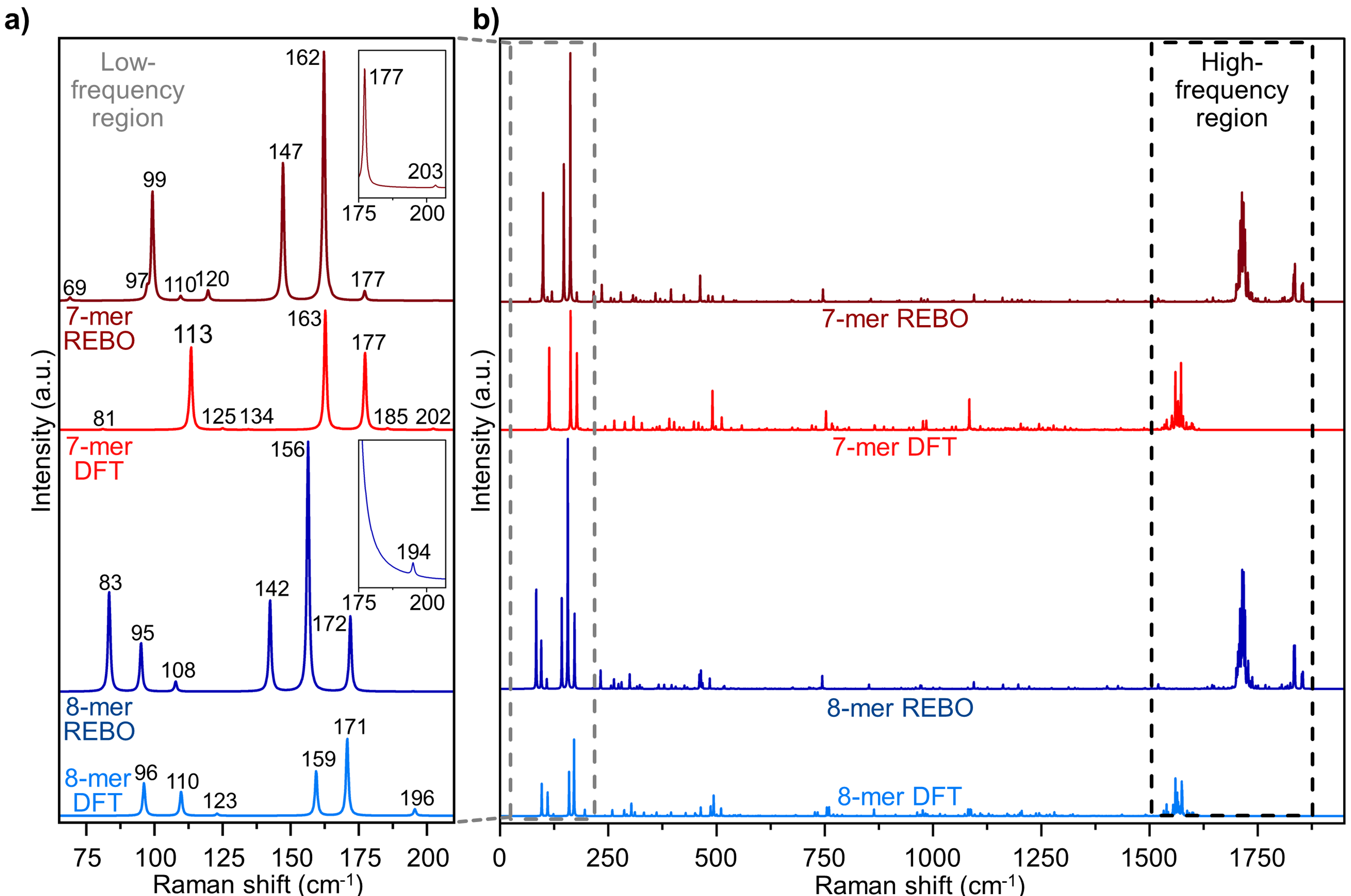}
    \caption{\textbf{Comparison of the peak positions calculated using DFT and REBOII}. a) Low-frequency region of the Raman spectra calculated using DFT for 7- (light red) and 8-mer (light blue) 17-AGNRs and REBOII (7-mer in dark red and 8-mer in dark blue). b) Full-range spectra calculated with the high-frequency region highlighted with a dashed square. The correspondence between REBOII and DFT is particularly good at low frequency.}
    \label{fS_DFT-vs-REBO}
\end{figure}
\FloatBarrier

\begin{figure}[!htbp]
    \centering
    \includegraphics[width=\textwidth]{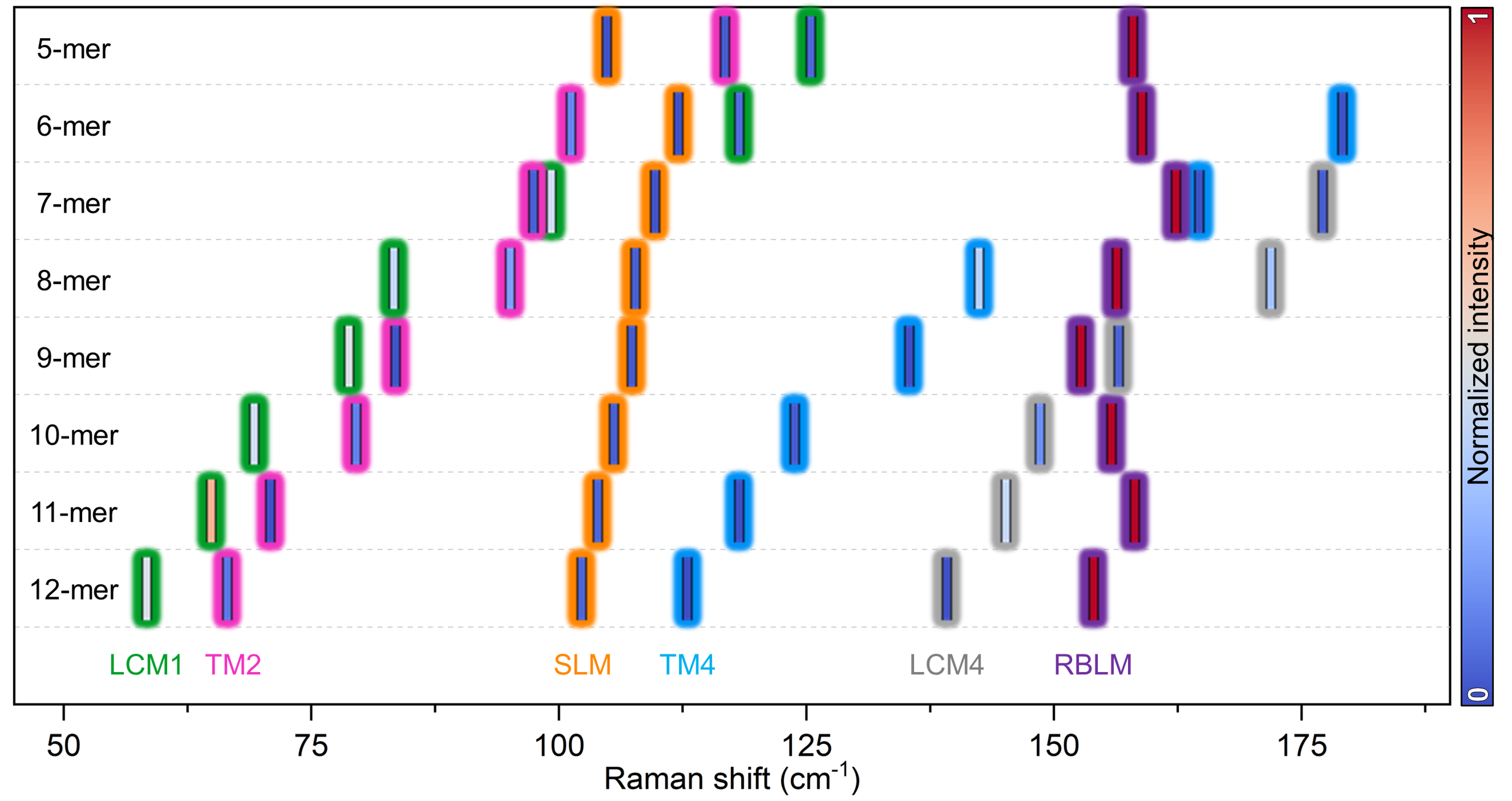}
    \caption{\textbf{The positions and intensities of the low-frequency modes calculated for 5- to 12-mer 17-AGNRs using REBOII}. Within each spectrum, the intensities are normalized to the most intense peak in the region (RBLM for all cases)  and are represented by the inner hue of the bars. The outer hues of the position bars represent different modes: LCM1 in green, TM2 in pink, SLM in orange, LCM4 in gray, and RBLM in violet. The numerals denote the (n-1)\textsuperscript{th} overtone of the modes.}
    \label{fS_rebo}
\end{figure}

A multitude of low-frequency modes such as LCMs, shear-like modes (SLMs), and in-plane twisting modes (TMs) are expected to appear at different positions. However, the relatively large length distribution experimentally observed for 17-AGNRs, a previous study on LCMs by Overbeck et al.~\cite{Overbeck2019Universal}, and our calculation suggest that the peaks appearing in this region are more likely to be LCMs. Hence, we assign the experimentally observed modes to the LCMs from ribbons of different lengths rather than contributions from other weaker vibrational modes.
Note that here we only discuss the effect of the length due to the discrepancies of the end structures of the ribbons. As previously discussed, experimentally, the GNR ends are not staggered (Figure~\ref{fS_End-segment}), which is not taken into account in the calculations.

\begin{figure}[!htbp]
    \centering
    \includegraphics[width=\textwidth]{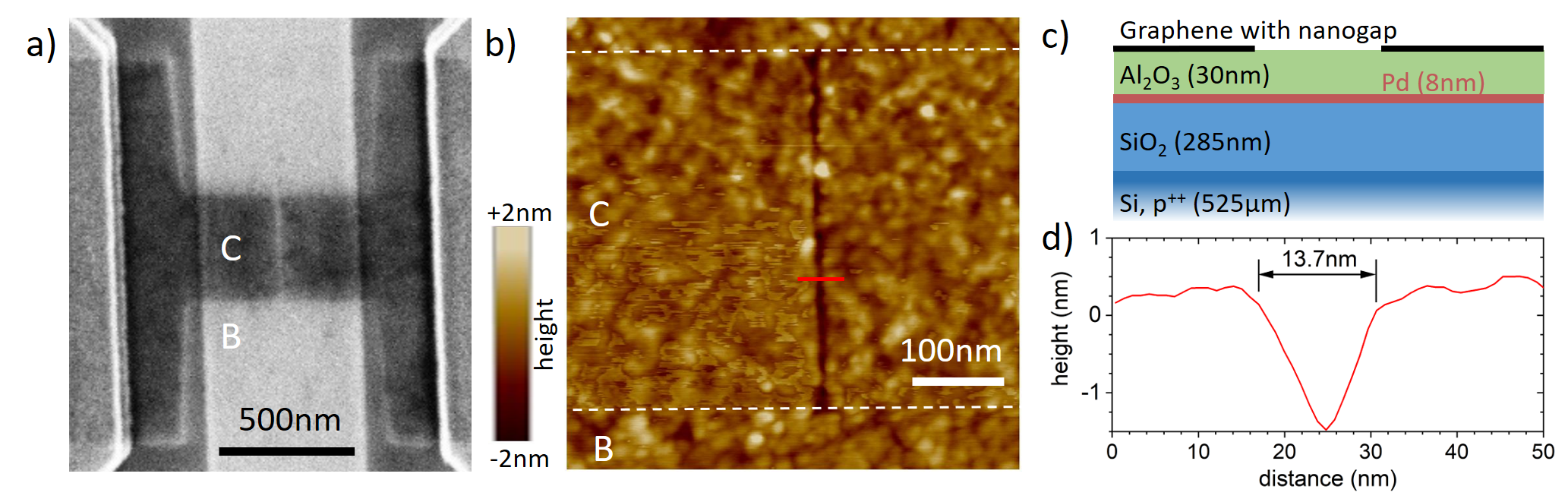}
    \caption{\textbf{Device and graphene electrode separation assessment by scanning electron microscope (SEM) and atomic force microscopy (AFM)}. a) SEM image of the device. Graphene outside the field of view is removed by a 1. EBL and RIE step. b) AFM scan of the central region of graphene electrodes. Different etching regions are labeled with B: 2. EBL and RIE step and C: graphene contact electrode. c) Schematic drawing of the vertical device structure. d) height profile at the red line in b). The graphene electrode separation is indicated.}
    \label{fS_device-assessment}
\end{figure}
\FloatBarrier

\begin{figure}[!htbp]
    \centering
    \includegraphics[width=\textwidth]{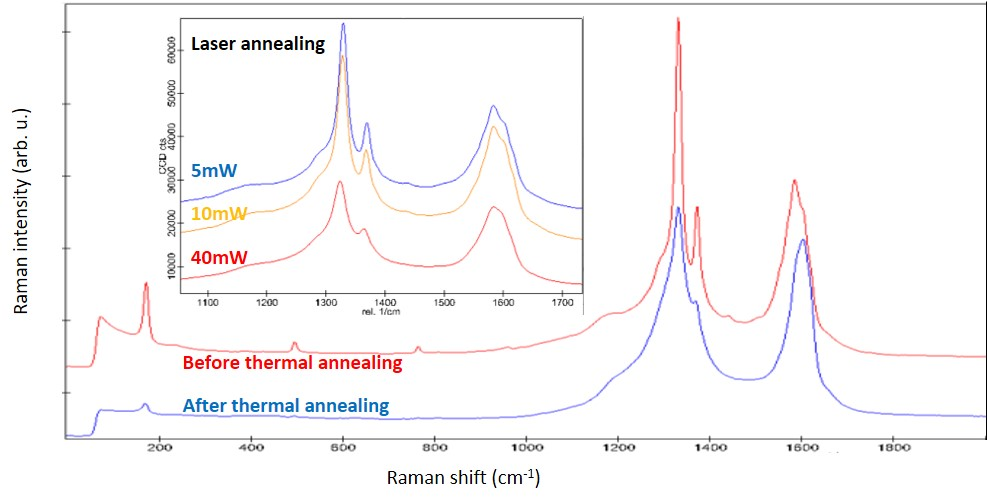}
    \caption{\textbf{Raman characterization of transferred 17-AGNRs on device substrates}. Visible is the clear separation of the peaks in the CH/D region (1200-1450 cm\textsuperscript{-1}) before thermal annealing (red). The peaks broaden after thermal annealing at 200~°C for 2 hours (blue). The inset shows Raman spectra acquired with different laser powers: 5 mW (blue), 10 mW (yellow), and 40 mW (red), leading to local thermal annealing and hence indicating a similar degradation.}
    \label{fS_Raman-on-device}
\end{figure}
\FloatBarrier

\begin{table}
\centering
\rotatebox{90}{
\begin{adjustbox}{max width=0.9\textheight}
\begin{tblr}{
  row{even} = {c}, 
  row{3,5,7,9,11,13} = {c},   
  cell{1}{1} = {c},
  cell{1}{3} = {c=15}{c}, 
  cell{2}{1} = {r=6}{},  
  cell{2}{2} = {r=2}{},
  cell{4}{2} = {r=2}{},
  cell{6}{2} = {r=2}{},
  cell{8}{1} = {r=6}{},
  cell{8}{2} = {r=2}{},
  cell{10}{2} = {r=2}{},
  cell{12}{2} = {r=2}{},
  vline{2-3} = {1}{}, 
  vline{2-4} = {2-13}{},
  vline{2-4} = {3,5,7,9,11,13}{},
  vline{2-4} = {4,6,10,12}{},
  hline{2,8} = {-}{},         
  hline{4,6,10,12} = {2-18}{},
}
Substrate &\textlambda\textsubscript{ex} (nm)& Peak information& &   &    &      &     &     &     &      &      &      &      &      &      &      &      \\
Au(111)   & 488              & Position    &    &     &    &      &     &     &     &      &      &      & 1302 & 1336 & 1370 & 1594 &    \\
          &                  & Intensity   &    &     &    &      &     &     &     &      &      &      & 194  & 632  & 316  & 836  &    \\
          & 532              & Position    &    &     &    &      &     &     &     &      &      & 1218 & 1305 & 1333 & 1375 & 1592 &    \\
          &                  & Intensity   &    &     &    &      &     &     &     &      &      & 132  & 324  & 984  & 328  & 1024 &    \\
          & 785              & Position    &    &     &    &      &     &     &     &      &      & 1205 & 1301 & 1330 &      & 1592 &    \\
          &                  & Intensity   &    &     &    &      &     &     &     &      &      & 35   & 59   & 69   &      & 121  &    \\
RO        & 488              & Position    & 68 & 103 & 120 & 169 &     &     &     & 1145 & 1176 &      &      & 1333 & 1370 & 1589 & 1603 \\
          &                  & Intensity   & 3  & 9   & 10  & 51  &     &     &     & 110  & 114  &      &      & 3959 & 1431 & 2893 & 2516 \\
          & 532              & Position    &    &     &    & 169  & 493 & 762 & 959 & 1144 &      &      & 1295 & 1333 & 1370 & 1589 & 1603 \\
          &                  & Intensity   &    &     &    & 142  & 29  & 22  & 8   & 34   &      &      & 201  & 911  & 353  & 562  & 459  \\
          & 785              & Position    &    &     &    &      &     &     &     &      &      & 1200 & 1282 &      & 1367 & 1587 &    \\
          &                  & Intensity   &    &     &    &      &     &     &     &      &      & 26   & 19   &      & 41   & 41   &    
\end{tblr}
\end{adjustbox}
}
  \caption{\textbf{Peak information of the experimental Raman spectra.} The units for the laser wavelengths and peak positions are~nm and cm\textsuperscript{-1}, respectively, and the unit for the intensity is arbitrary.} 
  \label{tS1}
\end{table}

\begin{table}
\centering
\rotatebox{90}{
\begin{adjustbox}{max width=0.92\textheight}
\begin{tblr}{
  cells = {c},
  cell{1}{3} = {c=21}{},
  cell{2}{1} = {r=8}{},
  cell{2}{2} = {r=4}{},
  cell{6}{2} = {r=4}{},
  cell{10}{1} = {r=8}{},
  cell{10}{2} = {r=4}{},
  cell{14}{2} = {r=4}{},
  vline{2-3} = {1}{},
  vline{2-4} = {2-17}{},
  vline{2-4} = {3-5,7-9,11-13,15-23}{},
  vline{2-4} = {6,14}{},
  hline{2,10} = {-}{},
  hline{6,14} = {2-23}{},
}
Substrate & \textlambda\textsubscript{ex} (nm)& Peak information& &   &     &       &     &     &     &      &      &      &      &       &        &       &      &      &       &       &       &       \\
Au(111)   & 488              & Position   &      &     &     &       &     &     &     &      &      &      &      & 1305  & 1336   & 1370  & 1393 &      & 1550  &       & 1593  & 1612  \\
          &                  & Intensity  &      &     &     &       &     &     &     &      &      &      &      & 106   & 563    & 211   & 176  &      & 73    &       & 754   & 233   \\
          &                  & FWHM       &      &     &     &       &     &     &     &      &      &      &      & 26    & 26     & 26    & 26   &      & 27    &       & 27    & 27    \\
          &                  & Area       &      &     &     &       &     &     &     &      &      &      &      & 4253  & 22559  & 8436  & 7049 &      & 3025  &       & 31146 & 9636  \\
          & 532              & Position   &      &     &     &       &     &     &     &      &      & 1219 &      & 1304  & 1334   & 1375  & 1410 & 1437 &       &       & 1593  &    \\
          &                  & Intensity  &      &     &     &       &     &     &     &      &      & 100  &      & 233   & 907    & 242   & 51   & 43   &       &       & 969   &     \\
          &                  & FWHM       &      &     &     &       &     &     &     &      &      & 30   &      & 19    & 19     & 19    & 19   & 19   &       &       & 25    &   \\
          &                  & Area       &      &     &     &       &     &     &     &      &      & 4740 &      & 6885  & 26836  & 7167  & 1514 & 1271 &       &       & 38666 &   \\
RO        & 488              & Position   & 67   & 104 & 120 & 170   &     &     &     & 1148 & 1176 &      &      & 1292  & 1333   & 1374  &      &      & 1550  & 1578  & 1592  & 1611  \\
          &                  & Intensity  & 7    & 14  & 11  & 53    &     &     &     & 88   & 73   &      &      & 325   & 3784   & 1101  &      &      & 547   & 1106  & 2071  & 1461  \\
          &                  & FWHM       & 6    & 4   & 13  & 18    &     &     &     & 23   & 23   &      &      & 23    & 23     & 23    &      &      & 24    & 24    & 24    & 24    \\
          &                  & Area       & 52   & 92  & 204 & 1443  &     &     &     & 3093 & 2563 &      &      & 11661 & 135632 & 39454 &      &      & 20413 & 41239 & 77173 & 54436 \\
          & 532              & Position   &      &     &     & 169   & 493 & 762 & 953 & 1141 & 1163 &      & 1277 & 1295  & 1329   & 1372  &      &      & 1545  & 1568  & 1586  & 1606  \\
          &                  & Intensity  &      &     &     & 150   & 27  & 20  & 3   & 30   & 19   &      & 72   & 112   & 850    & 253   &      &      & 161   & 276   & 437   & 352 \\
          &                  & FWHM       &      &     &     & 10    & 10  & 10  & 10  & 26   & 26   &      & 21   & 21    & 21     & 21    &      &      & 21    & 21    & 21    & 21   \\
          &                  & Area       &      &     &     & 2418  & 442 & 331 & 51  & 1232 & 781  &      & 2306 & 3625  & 27383  & 8167  &      &      & 5179  & 8881  & 14047 & 11331   
\end{tblr}
\end{adjustbox}
}
  \caption{\textbf{Peak information of the experimental Raman spectra after deconvolution.} The $R^{2}$ values of the deconvolutions for the spectra measured on Au(111) at \textlambda\textsubscript{ex} = 488 and 532~nm are 0.88 and 0.97, respectively, and 0.99 for those measured on RO. The positions and FWHMs are in cm\textsuperscript{-1}, intensities are in arbitrary units (a.u.), and the areas are in a.u.$\cdot$cm\textsuperscript{-1}.} 
  \label{tS2}
\end{table}

\begin{table}
\centering
\begin{tblr}{
  cells = {c},
  hline{2} = {-}{},
}
17-AGNR & Length (nm) \\
5-mer   & 2.414       \\
6-mer   & 2.840       \\
7-mer   & 3.266       \\
8-mer   & 3.692       \\
9-mer   & 4.118       \\
10-mer  & 4.544       \\
11-mer  & 4.970       \\
12-mer  & 5.396
\end{tblr}
\caption{\textbf{The lengths of the short 17-AGNRs simulated using REBOII.}}
\label{tS3}
\end{table}
\FloatBarrier

\end{document}